\documentclass[twocolumn,showpacs,preprintnumbers,prb,amsmath,amssymb]{revtex4}
\usepackage{graphicx}
\usepackage{dcolumn}
\usepackage{bm}
\usepackage{textcomp}
\usepackage{amsmath,amssymb}
\usepackage{epsf}
\usepackage{graphicx}
\usepackage{epsfig}
\usepackage{subfigure}
\begin{document}
\title{Slow dynamics of interacting antiferromagnetic nanoparticles}
\author{Sunil K. Mishra and V. Subrahmanyam}
\address{Department of Physics, Indian Institute of Technology Kanpur-208016, India}
\date{\today}
%\ead{sunilkm@iitk.ac.in}
\begin{abstract}
We study magnetic relaxation dynamics, memory and aging effects in interacting polydisperse antiferromagnetic NiO nanoparticles by solving a master equation using a two-state model. We investigate the effects of interactions using dipolar, Nearest-Neighbour Short-Range (NNSR) and Long-Range Mean-Field (LRMF) interactions. The magnetic relaxation of the nanoparticles in a time-dependent magnetic field has been studied using LRMF interaction. The size-dependent effects are suppressed in the ac-susceptibility, as the frequency is increased. We find that the memory dip, that quantifies the memory effect is about the same as that of non-interacting nanoparticles for the NNSR case. There is a stronger memory-dip for LRMF, and a weaker memory-dip for the dipolar interactions. We have also shown a memory effect in the Zero-field-cooled magnetization for the dipolar case, a signature of glassy behaviour, from Monte-Carlo studies. 
\end{abstract}
\pacs {75.50.Ee, 75.50.Tt, 75.75.Fk, 75.75.Jn} 

\maketitle
%%%%%%%%%%%%%%%%%%%%%%%%%%%%%%%%%%%%%%%%%%%%%%%%%%%%%%%%%%%%%%%%%%%%%%%%%%%
\section{Introduction}
%%%%%%%%%%%%%%%%%%%%%%%%%%%%%%%%%%%%%%%%%%%%%%%%%%%%%%%%%%%%%%%%%%%%%%%%%%%
%From the recent years study of slow dynamics has become important in characterizing the assembly of the nanoparticles.  
%The study of magnetism in nanoparticles has been
%Magnetic nanoparticles are of great interest in the recent decade
%The interest in the study of magnetism in nanoparticles has been renewed 
%from the last few decades
Magnetism in nanoparticles has received enormous attention in recent years due to its
 technological \cite{weller,richter,berry}
 as well as fundamental research aspects. \cite{fiorani, kodama,kodama1,jonsson1,jonsson2,sahoo1,dormann,labarata,jonsson3,djuberg,malay2,garcia1,jonsson,mamiya,sahoo2,sun,sahoo3,sahoo4,zheng,sasaki,tsoi,malay,wang,du,suzuki,wang1,winkler,makhlouf,tiwari}.
Amid many studies related to magnetic nanoparticles, the important concerns were related to the relaxation behaviour of the assemblies of nanoparticles which has been addressed in the recent times. \cite{jonsson1,sahoo1,dormann,jonsson2,jonsson3,labarata}
The dynamics of an assembly of nanoparticles at low temperatures has gained a 
lot of attention over the last few years. In a dilute system of nanoparticles,
 the interparticle interaction is very small as compared to the anisotropy energy
 of the individual particles. These isolated particles follow the dynamics in accordance 
with the N\'eel-Brown model, \cite{brown} and the system is known as superparamagnetic. The giant spin moment of the nanoparticles
thermally fluctuates between their easy directions at high temperatures.
 As the temperature is lowered towards a blocking temperature, 
the relaxation time becomes equal to the measuring time and the superspin moments freeze 
along one of their easy directions. As the role of interparticle interaction becomes significant,
 the nanoparticles do not behave like individual particles; rather, their dynamics is governed by the 
collective behaviour of the particles, like in a spin glass. \cite{jonsson1,jonsson2,sahoo1,dormann,labarata,jonsson3} 

In our recent work, \cite{sunil1} we studied the slow dynamics of NiO nanoparticles distributed sparsely so that particle-particle interactions could be neglected. However, in reality, interparticle interactions play  a major role in describing various interesting phenomena observed experimentally in a collection of nanoparticles. This leads us to include interactions among the particles in our study. 
If the interparticle interaction is too small, the dynamics of an assembly of nanoparticles is a result of the individual nanoparticle dynamics. 
Increasing the particle-particle interaction, the behaviour of the
system becomes more complicated, even if we consider each nanoparticle as a single giant-spin. The dynamics of the assembly of nanoparticles involves various degrees of freedom related to the individual particles coupled to each other with interparticle interactions. 
In this paper, we will discuss various interactions comparatively.

The main type of magnetic interactions that can play an important role 
in the assemblies of nanoparticles are: (i) long-range interactions (for example, dipolar interactions) among the particles, and (ii) short-range exchange-interactions arising due to the
surface spins of the particles which are in close contact.
The dipolar interaction is a peculiar kind of interaction which may
favour ferromagnetic or antiferromagnetic alignment of the
magnetic moments depending upon the geometry of the system. 
This interaction among the nanoparticles may give rise to a collective behaviour which may lead to a dynamics similar to that of a spin-glass. A resemblance from spin glass system owes to a random distribution of easy axes, which causes disorder and frustration of
the magnetic interactions. The complex interplay between the
 disorder and frustration determines the state of the
system and its dynamical properties. These systems are widely known as superspin glasses. This superspin glass phase has been characterised by observations of a critical 
slowing-down, \cite{djuberg} a divergence in
the non-linear susceptibility, \cite{jonsson} and aging and relaxation effects in the low-frequency ac susceptibility. \cite{mamiya}
The Monte Carlo simulations on the system of an assembly of nanoparticles show aging \cite{andersson} and magnetic 
relaxation behaviour \cite{ulrich} like in a spin glass. %but simulations of zero field cooling (ZFC) and 
%field cooling (FC) susceptibilities show no indication of spin-glass ordering \cite{garcia}.
For an interacting assembly of nanoparticles, the aging
 and memory effects are the two aspects that have been studied extensively in recent years, however, mostly in the case of
 ferromagnetic particles. \cite{jonsson1,jonsson2,sahoo1,jonsson,sahoo2,sun,sahoo3,sahoo4,zheng,sasaki,tsoi,malay,wang,du,suzuki,wang1}

In this paper, we investigate the effects of interactions on a collection of a few antiferromagnetic nanoparticles, which has received relatively lesser attention. Recently, we have shown that for NiO nanoparticles, a combined effect of the
 surface-roughness effect and finite-size effects in the core magnetization leads to size-dependent fluctuations in net magnetic-moment. \cite{sunil} 
These size-dependent fluctuations in the magnetization lead to a dynamics which is qualitatively different from that of ferromagnetic nanoparticles.
We will study the dynamics of the system by analytically solving the master equation for a two-state model.
We will invoke various interactions and 
study the relaxation phenomena under these interactions comparatively.
We will show the effect of the size-dependent magnetization fluctuations 
on the time-dependent properties of an interacting assembly 
of nanoparticles. We will also compare the dynamics with that of the non-interacting case. 
 We will discuss the  memory effects in the FC as well as the ZFC magnetization protocols. The organisation of this paper is as follows. 
In Section \ref{mod}, we discuss the two-state model. We discuss the ZFC and FC magnetizations in Section \ref{seczfc_int}, and the ac-susceptibility study in Section \ref{secac}. In Section \ref{secmemory_int}, we show the memory effect investigations. The aging effect has been presented in Section \ref{secaging_int}. Finally, we summarise in Section \ref{secconclusion_int}.
%%%%%%%%%%%%%%%%%%%%%%%%%%%%%%%%%%%%%%%%%%%%%%%%%%%%%%%%%%%%%%%%%%%%%%%
\section{ Relaxation phenomena and various interactions}
\label{mod}
%%%%%%%%%%%%%%%%%%%%%%%%%%%%%%%%%%%%%%%%%%%%%%%%%%%%%%%%%%%%%%%%%%%%%%%%%
\begin{figure}[t]
%\vspace*{1.2cm}
\begin{center}
\includegraphics [angle=0,width=1.0\linewidth] {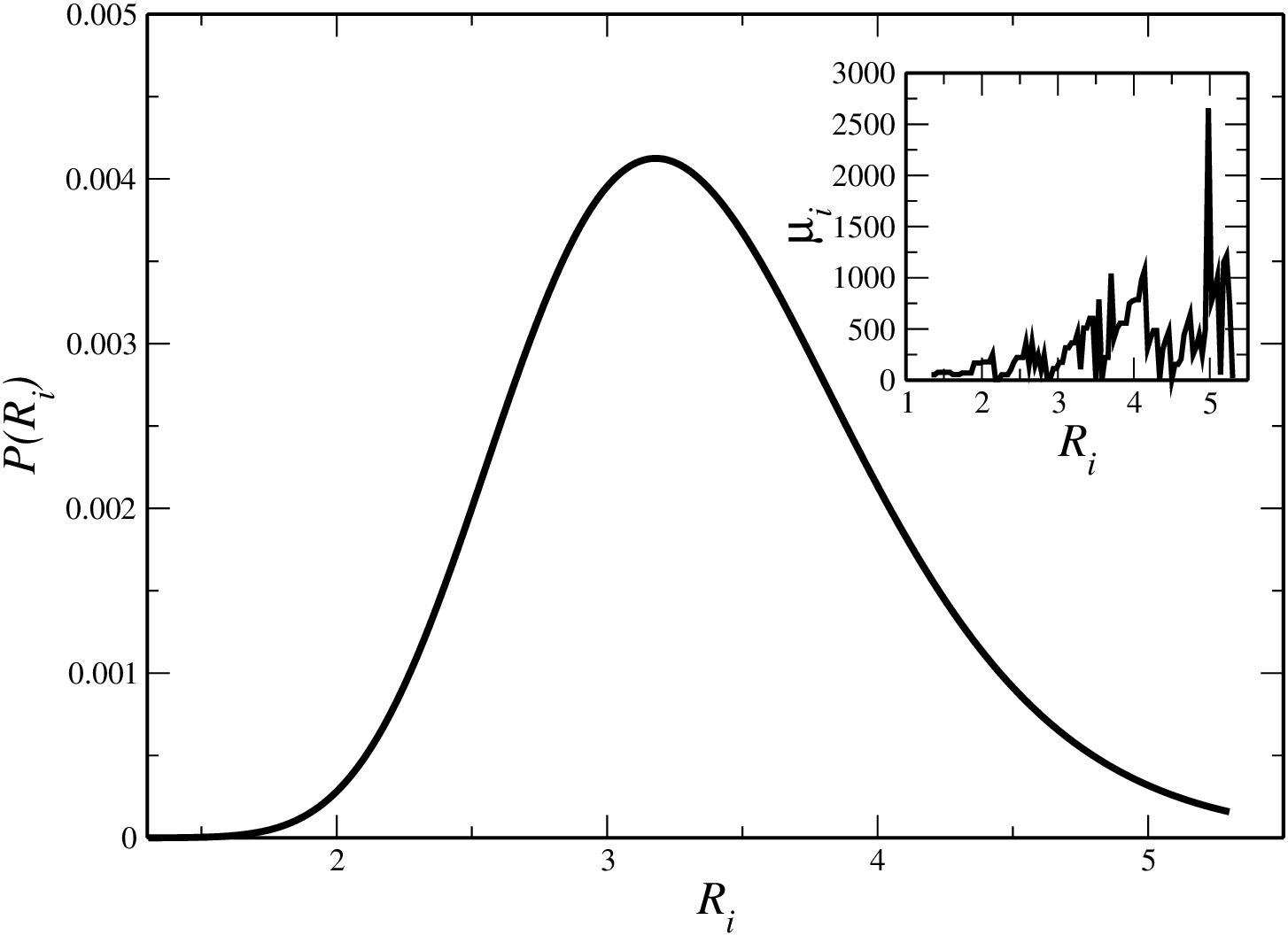}
\caption{ A lognormal distribution of nanoparticles with sizes ranging from $1.3a_0$-$5.3a_0$, where $a_0=4.17\AA$. The width of the distribution is $0.6$. The inset displays the magnetic moment vs the particle size for the same size range. The magnetic moment has a non-monotonic and oscillatory dependence on $R$.}
\label{fig1a}
\end{center}
\end{figure}
We assume a nanoparticle to be a giant spin in which the exchange interactions are so strong that all the spins in the particle show a coherent rotation in unison. Hence, the dynamics of a system of nanoparticles can be easily understood as the dynamics of a giant spin-moment. Here, we are dealing with antiferromagnetic nanoparticles, which display size dependent fluctuations in the magnetization  as shown in the inset of the figure FIG. \ref{fig1a}.
On the other hand, the net magnetic moment  of ferromagnetic nanoparticles shows a linear dependence on the volume of the particles. \cite{sasaki} As we study the dynamics of antiferromagnetic nanoparticles, we 
%In the present study for the AFN, the magnetic moment does not depend linearly on size, instead it shows size dependent fluctuations.
 might expect the role of these size-dependent fluctuations in the magnetization to be manifested in the time-dependent properties of these antiferromagnetic nanoparticles.
We use a simple model where the energy of each particle $i$
is contributed by the anisotropy energy and the
 Zeeman energy. Now, in order to incorporate interactions in the present model, we add an interaction energy term $\mathcal{V}_{ij}$ in the Hamiltonian. Thus, the Hamiltonian is written as:
\begin{eqnarray}
\mathcal{H}=-\sum_i\mathcal{K}V_i-\sum_i\vec{\mu}_i.\vec{H}+ \sum_i\sum_{j\neq i} \mathcal{V}_{ij},
\label{modl}
\end{eqnarray}
% $\mathcal{M} $
and the energy of a particle is given by
\begin{eqnarray}
E_i=-\mathcal{K}V_i-\mu_i H+ \sum_{j\neq i} \mathcal{V}_{ij},
\label{modl}
\end{eqnarray}
where $\vec{\mu}_i$ is the magnetic moment of the $i^{\rm th}$ particle, $\mu_i=\arrowvert\vec{\mu}_i\arrowvert$, $\mathcal{K}$ is the anisotropy constant, %and $\mathcal{H}$ is the applied magnetic field
 and $H$ is the applied field. Because of the interaction term  $\mathcal{V}_{ij}$, the energy of the $i^{\rm th}$ particle depends on the state of all the other particles. We assume each nanoparticle to be an ising spin, and the magnetic moment of each particle to be $\vec{\mu}_i=\mu_is_i\hat{z}$, with $s_i=\pm1$.
The various interactions used in the present study are: 
\begin{enumerate}
 \item [i)]  A Nearest-Neighbour Short-Range (NNSR) type interaction, which can be written as:
\begin{eqnarray}
 \mathcal{V}_{ij}^{\rm{sr}}=J^{\rm{sr}}\sum_{<i,j>}\mu_i \mu_j,
\label{def_sr}
\end{eqnarray}
where the summation is taken over the nearest-neighbour sites. Here $J^{\rm{sr}}$ is the NNSR interaction strength.
\item[ii)] A long-Range, Mean-Field (LRMF) type interaction,\cite{gupta} which is given as:
\begin{eqnarray}
 \mathcal{V}_{ij}^{\rm{lr}}&=&\frac{J^{\rm{lr}}}{2N}(\sum_{i=1}^N \mu_i)^2 \nonumber \\
&=&\frac{J^{lr}}{2N} \sum_{i=1}^N\mu_i^2+\frac{J^{lr}}{N}\sum_{j\neq i}\mu_i\mu_j,
\label{intlr_def}
\end{eqnarray}
where $N$ is the total number of nanoparticles and $J^{\rm{lr}}$ is interaction parameter. The above equation Eq. (\ref{intlr_def}) is scaled by $N$ in order to insure that the total interaction-energy is proportional to $N$ and not $N^2$. 
\item [iii)] A long-range dipolar interaction, with  the particle $j$ separated by a distance $r_{ij}$ from particle $i$, given as:
\begin{eqnarray}
\mathcal{V}_{ij}^{\rm{dipolar}}=\alpha \sum_{j\neq i} \left[ \frac{\vec{\mu_i} \vec{\mu_j}}{r_{ij}^3} -\frac{3\left( 
\vec{\mu_i}.\vec{r_{ij}}\right) \left( \vec{\mu_j}.\vec{r_{ij}}\right) }{r_{ij}^5}\right], 
\label{intdip_def}
\end{eqnarray}
where $\alpha=\mu_0 \mu_B^2/4\pi r_0^2$ characterises the strength of the dipolar interaction and $r_0$ is the lattice parameter of the lattice, where nanoparticles are placed. Variations in $r_0$ may result the nanoparticles to be getting closer or away from each other and hence, increasing or decreasing the effects of dipolar interactions.
%We consider $\alpha=5\times10^{-7}$ for interacting cases discussed below.
\end{enumerate}
Let us consider a model case of NNSR. Using the mean-field approximation, %as discussed in Section \ref{mftheory}, 
we can find a mean field felt by the $i^{\rm th}$ particle due to interactions with its neighbours. We can replace Eq. (\ref{def_sr}) by its mean-field form as:
\begin{eqnarray}
 \mathcal{V}_{ij}^{\rm MF}&=&\mu_i\sum_{j\neq i}\langle \mu_j\rangle, \nonumber \\
&=& \mu_i H_i^{\rm MF},
\end{eqnarray}
where $H_i^{\rm MF}$ is the mean field on the $i^{\rm th}$ spin from all the nearest neighbour spins. Thus, in this framework, each particle can be viewed under the influence of a local magnetic-field:
 \begin{eqnarray}
  H_i^{\rm eff}=H +H_i^{\rm MF}.
 \end{eqnarray}
The effect of interactions on a nanoparticle can be incorporated by replacing the external field by an effective field which may lead to a mean-field equation of state and can be solved self-consistently. 
Hereafter, in all the cases of interactions, we will define an effective field $H_i^{\rm{eff}}$ on a nanoparticle $i$ as the sum of the external field $\mathcal{H}$ and a locally-changing interaction-field i.e.,
\begin{eqnarray}
H_i^{\rm{eff}}=H + \frac{1}{\mu_i}\sum_{j\neq}\mathcal{V}_{ij}.
\label{heff_int} 
\end{eqnarray}  
The above equation shows that the role of the interactions is to modify the energy barrier, which is solely due to the anisotropy 
contributions of each particle in the non-interacting case. For strong interactions, their effects become dominant and
the individual energy-barriers can no longer be considered to be the only relevant energy-scale.
In this
case, the relaxation is governed by a co-operative phenomenon of the system. 
The energy landscape with a complex hierarchy of local
minima is similar to that of a spin glass. We should note that in contrast with the static energy barrier distribution
arising only from the anisotropy contribution, the reversal of
one particle moment may change the energy barriers of the
assembly, even in the weak-interaction limit. Therefore, the
energy-barrier distribution gets modified as the magnetization
relaxes.

 By defining an effective field given by Eq. (\ref{heff_int}), we can map an interacting assembly of nanoparticles as a collection of non-interacting nanoparticles, where each particle experiencesan effective field $H_i^{\rm{eff}}$ which gets modified as the temperature changes.
In the absence of the magnetic field, the
superparamagnetic relaxation-time for the thermal activation over the energy barrier $\mathcal{K}V_i$ is given by $ \tau_i =\tau_0 \exp (\mathcal{K}V_i/k_B T)$,
  where $\tau_0$, the microscopic time, is of the order of $10^{-9}$ s.  The anisotropy 
constant $\mathcal{K}$ has a typical value of about $4\times10^{-1}$ J $\rm{cm}^{-3}$ for NiO. \cite{hutchings}

  The occupation probabilities with the magnetic moments parallel and antiparallel
to the magnetic field direction are denoted by $p_1(t)$ and $p_2 (t) = 1 - p_1(t)$,
respectively. The magnetic moment of each particle is supposed to occupy one of the two available states with energies $-\mathcal{K}V_i+\mu_i^{\rm s}H_i^{\rm{eff}}$ or $-\mathcal{K}V_i-\mu_i^{\rm{s}}H_i^{\rm{eff}}$, where $\mu_i^s$ is the saturation magnetic-moment of the $i^{\rm th}$ particle. These probabilities satisfy a
master equation, which is given by,
\begin{eqnarray}
  \frac{d}{dt}p_1(t)=-\frac{1}{\tau_i}p_1(t)+\frac{1}{2\tau_i}\left[  1+\frac{{\mu}_i^{\rm s} H_i^{\rm eff}}{T}\right]. 
\label{mseqn2}
\end{eqnarray}
The magnetic moment of each particle of volume $V_i$ can be written as ${\mu_i} = \left[ 2p_1 - 1\right]  \mu_i^s.$ as 
\begin{eqnarray}
 \frac{d}{dt}{\mu_i}=-\frac{\mu_i}{\tau_i}+\frac{1}{2\tau_i} \left[1+\frac{\left( {\mu}_i^{\rm s}\right)^2  H_i^{\rm eff}}{T} \right] . 
\label{int_muexp}
\end{eqnarray}
We can define magnetization of each particle as $M_i\equiv\mu_i/V_i$. Since $H_i^{\rm eff}$ also contains a summation over all the other $\mu_j$'s interacting with the particle, the right-hand side of the above equation can be written as the sum of the interaction term with $j^{\rm th}$ particle and the term containing the magnetic field. Thus, we can write the Eq. (\ref{int_muexp}) as: 
\begin{eqnarray}
 \frac{d}{dt} M_i = A(i,j)M_j-a_{0i}, 
\label{component_eqn}
\end{eqnarray}
Here $A(i,j)$ represents the interaction term which is given below for each case separately, and $a_{0i}=-\left( {\mu}_i^{\rm s}\right)^2 H/V_i\tau_i T$.
\begin{enumerate}
 \item [(i)] For NNSR interaction, $A(i,j)$ is given as:
\[ A(i,j) = \left\{ \begin{array}{ll}
          -\frac{1}{\tau_i}, &\mbox{if $i = j$};\\
        \frac{J^{\rm sr}\left( {\mu}_i^{\rm s}\right)^2  }{V_i \tau_i T} \delta_{\vec{r},\vec{r}+\epsilon}, & \mbox{if $i \neq j$}\end{array} \right. \]
where $\vec{r}$ is the position vector of any site and $\epsilon$ is the nearest-neighbour distance.  
\item [(ii)] For LRMF interaction, we can write $A(i,j)$ as 
\[ A(i,j) = \left\{ \begin{array}{ll}
          -\frac{1}{\tau_i}+\frac{J^{\rm lr}}{2N}\frac{\left( {\mu}_i^{\rm s}\right)^2 }{V_i\tau_i T}, &\mbox{if $i = j$};\\
        \frac{J^{\rm{lr}}}{N}\frac{\left( {\mu}_i^{\rm s}\right)^2}{V_i\tau_i T}, & \mbox{if $i \neq j$}.\end{array} \right. \] 
\item [(iii)] For dipolar interaction we can write,
\[ A(i,j) = \left \{ \begin{array}{ll}
          -\frac{1}{\tau_i}, &\mbox{if $i = j$};\\
        \frac{\alpha \left( {\mu}_i^{\rm s}\right)^2}{V_i \tau_i T} \frac{\left( r_{ij}^2-3z_{ij}^2\right) }{r_{ij}^5}, & \mbox{if $i \neq j$}.\end{array} \right. \]
\end{enumerate}
There are $N$ equations of the form given by Eq. (\ref{component_eqn}) for the system of $N$ particles.  We can write these N equations in a matrix form as:
\begin{eqnarray}
  \frac{d}{dt}\hat{\mathcal{M}} = \mathcal{A}\hat{\mathcal{M}}-\hat{a}_0, 
\label{mateq}
\end{eqnarray}
where $\mathcal{A}$ is a $N\times N$ matrix whose elements are given by $A(i,j)$, defined above, and $\hat{\mathcal{M}}$ and $\hat{a}_0$ are vectors of length $N$. The elements of $\hat{\mathcal{M}}$ are the $M_i$'s, and those of $\hat{a}_0$ are the $a_{0i}$'s.
Eq. (\ref{mateq}) can be solved for the various interactions defined above.
The formal solution of the matrix equation Eq. (\ref{mateq}) takes the form:
\begin{eqnarray}
 \hat{\mathcal{M}}(t)=e^{At}\hat{C}+\hat{\nu},
\label{trial_sol}
\end{eqnarray}
where $\hat{\nu}=\mathcal{A}^{-1}\hat{a}_0$, a time-independent solution of Eq. (\ref{trial_sol}) and $\hat{C}=\hat{\mathcal{M}}(0)-\hat{\nu}$, $\hat{\mathcal{M}}(0)$ being the value of  $\hat{\mathcal{M}}$ at $t=0$. Thus we can write,
\begin{eqnarray}
 \hat{\mathcal{M}}(t)=e^{\mathcal{A}t}\hat{\mathcal{M}}(0)+(1-e^{\mathcal{A}t})\hat{\nu}.
\label{mmu}
\end{eqnarray} 
The matrix $\mathcal{A}$ in the above equation Eq. (\ref{mmu}) is a general matrix of order $N\times N$. To evaluate $e^{\mathcal{A}t}$, we use  the MATLAB function {\it expm} which implements a diagonal Pad\'{e} approximation of exponential of the matrix with a scaling and squaring technique. \cite{golub,higham}

 Pad\'{e} approximation to an exponential of a matrix $\mathcal{B}$ is written as:
\begin{eqnarray}
 e^{\mathcal{B}}\approx R_{pq}(\mathcal{B})= \left[ D_{pq}(\mathcal{B})\right]^{-1} N_{pq}(\mathcal{B}) 
\end{eqnarray}
where the numerator term is given as
\begin{eqnarray}
 N_{pq}(\mathcal{B})=\sum_{j=0}^p\frac{(p+q-j)!p!}{(p+q)!j!(p-j)!}{\mathcal{B}}^{j}
\label{nume_pade}
\end{eqnarray}
and the denominator term is written as
\begin{eqnarray}
 D_{pq}(\mathcal{B})=\sum_{j=0}^q\frac{(p+q-j)!q!}{(p+q)!j!(q-j)!}(-\mathcal{B})^{j}
\label{denom_pade}
\end{eqnarray}
Diagonal Pad\'{e} approximants can be calculated by taking $q=p$ in Eqs. (\ref{nume_pade}) \&  (\ref{denom_pade}). The index $p$ depends upon the norm of the matrix $\mathcal{B}$. Using a backward-error analysis, $p$ can be calculated. \cite{higham}
Unfortunately, the Pad\'{e} approximants are accurate only near the origin. This problem can be overcome by the scaling and squaring technique which exploits,
\begin{eqnarray}
 e^{\mathcal{B}} =\left( e^{\mathcal{B}/2^k}\right)^{2^k}. 
\end{eqnarray}
Firstly, a sufficiently large $k$ is chosen such that $\mathcal{B}/2^k$ is close to the zero matrix. Then, a diagonal Pad\'{e} approximant is used to calculate $\exp(\mathcal{B}/2^k)$. Finally, the result is squared $k$ times to obtain the approximation to $e^{\mathcal{B}}$. 
We can use Pad\'{e} approximation with a scaling and squaring technique to calculate $e^{At}$ in Eq. (\ref{mmu}). However, the norm of the matrix $\mathcal{A}$, $\|\mathcal{A}\|\approx10^9$, is too large in all the three cases of interactions in Eqs. (\ref{def_sr}), (\ref{intlr_def}) \& (\ref{intdip_def}), for $N=1000$. The scaling and squaring method comes very handy to overcome this large value of the norm. \cite{higham}  In this case, the degree of the approximant is $p=13$. 
The total magnetic moment of the system of nanoparticles with volume distribution $P(V_i)$ is given by
\begin{eqnarray}
%\begin{eqnarray}
 \bar{M}=\frac{1}{N}\sum_{i=1}^N M_i=\int {M}_i P(V_i) d V_i.
\label{mtvpv_int}
%\end{eqnarray}
% \bar{M}=\int {M}_i P(V_i) d V_i.
%\label{mtvpv_int}
\end{eqnarray}

The size-distribution plays a significant role in the overall dynamics of the system of nanoparticles  governed by equation (\ref{mtvpv_int}). The exponential dependence of $\tau$ on the particle size $V_i$ assures that even a weak polydispersity may lead to a broad distribution of relaxation times, which gives rise to an interesting slow dynamics. 
For a dc measurement, if relaxation time coincides with the measurement time scale $\tau_m$, we can define \cite{bean} a critical volume $V_{\rm{B}}$ as $\mathcal{K}V_{\rm{B}}=k_{\rm{B}} T_{\rm{B}} \rm{ln}(\tau_m/\tau_0)$, where $T_{\rm{B}}$ is referred as blocking temperature.
The critical volume $V_{\rm{B}}$ has a strong linear dependence on $T_B$ and weakly logarithmic dependence on the observation time scale $\tau_m$. If the volume of the particle $V_i$ in a 
polydisperse system is less than $V_{\rm{B}}$, the super spin would have undergone many rotations within the measurement time scale with an average magnetic moment zero. These particles are termed as {\it superparamagnetic} particles. On the other hand if $V_i > V_{\rm{B}}$, the super spins can not completely rotate within the measurement time window and show {\it blocked or frozen} behavior. However, the particles having volume $V_i\simeq V_{\rm{B}}$ are in {\it dynamically active} regime.
\begin{figure}
%\hspace*{-.8cm}
\centering
%\centering
\begin{tabular}{c}
\epsfig{file=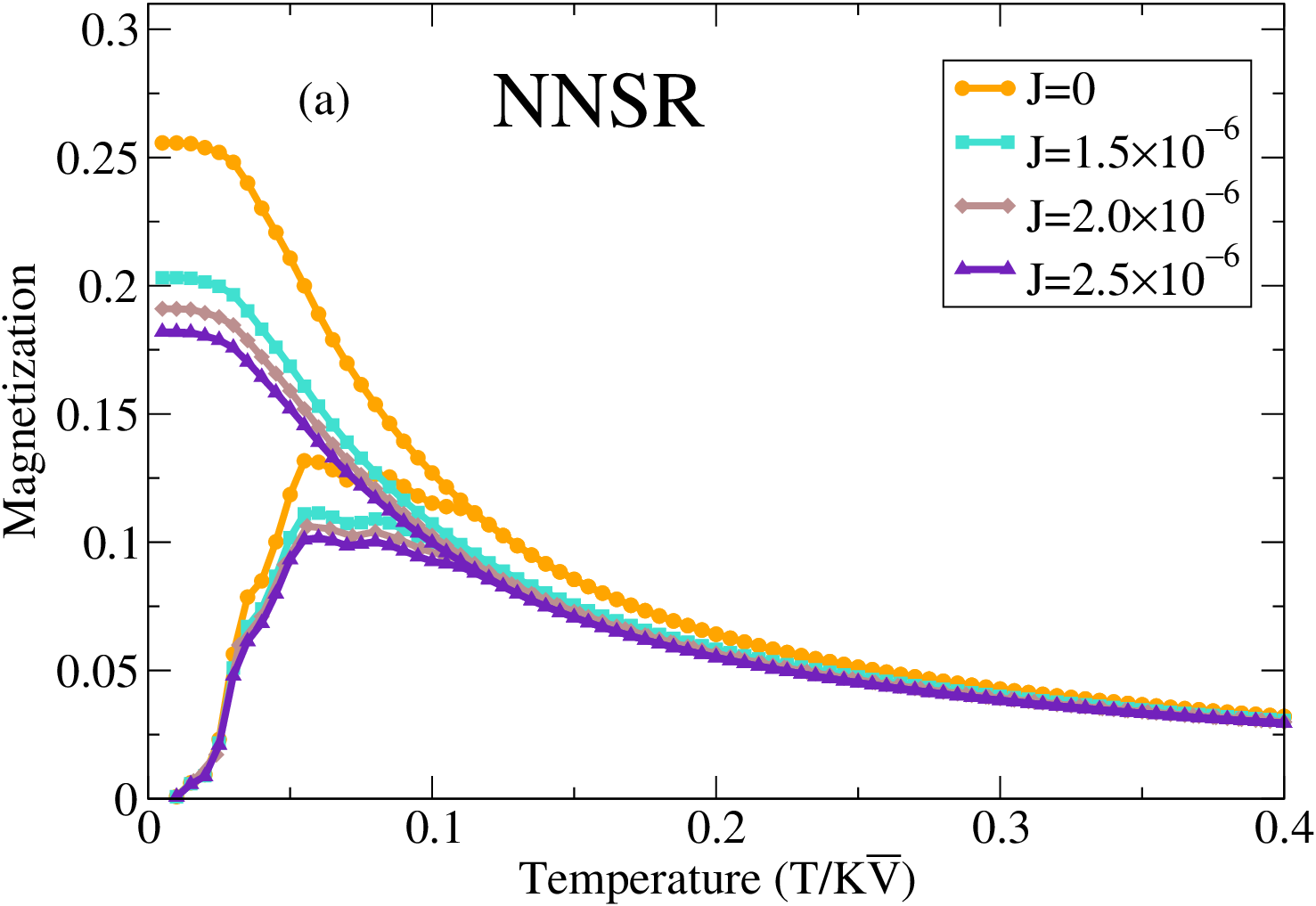,width=1.0\linewidth,clip=} \\
\epsfig{file=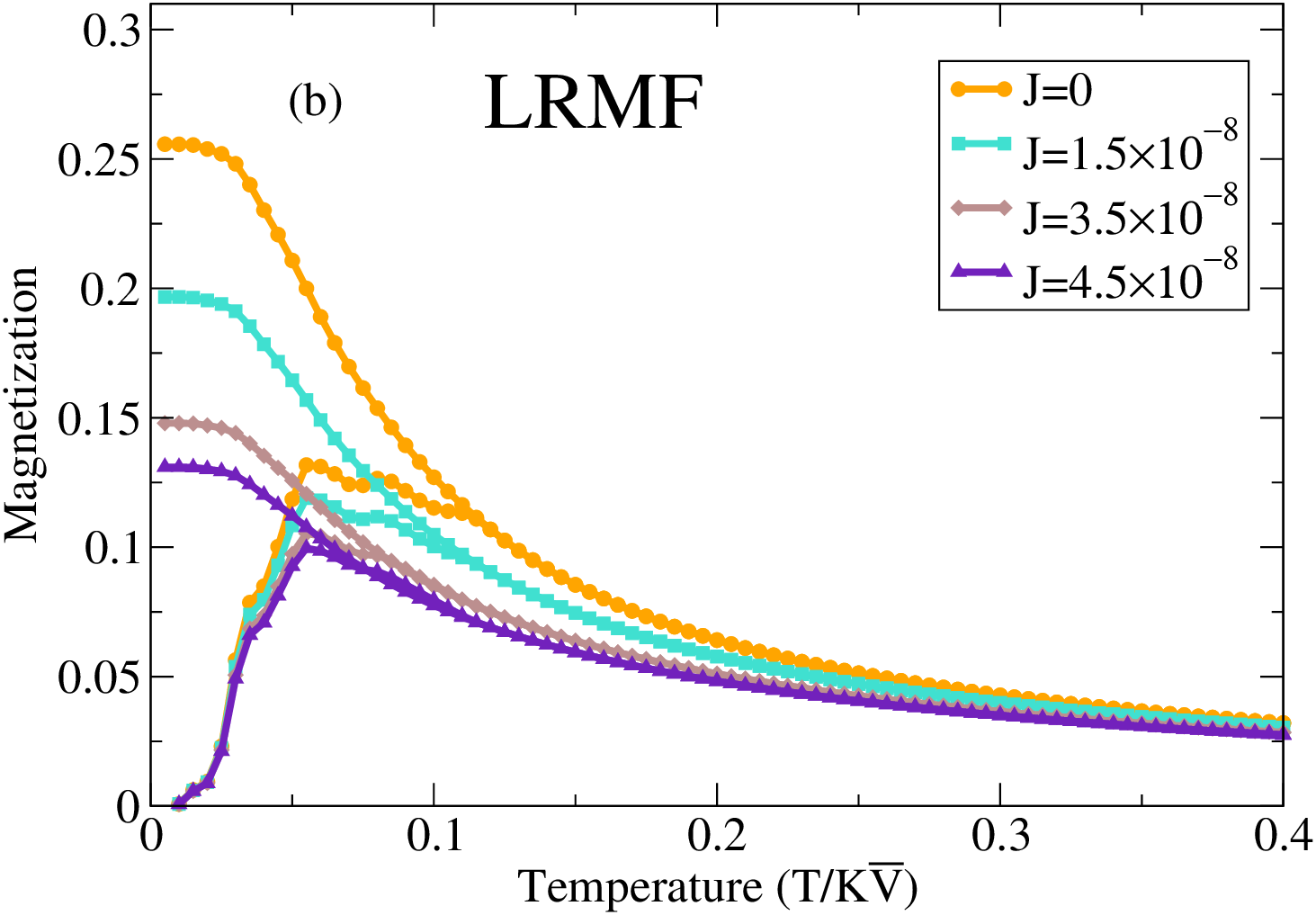,width=1.0\linewidth,clip=} \\
\epsfig{file=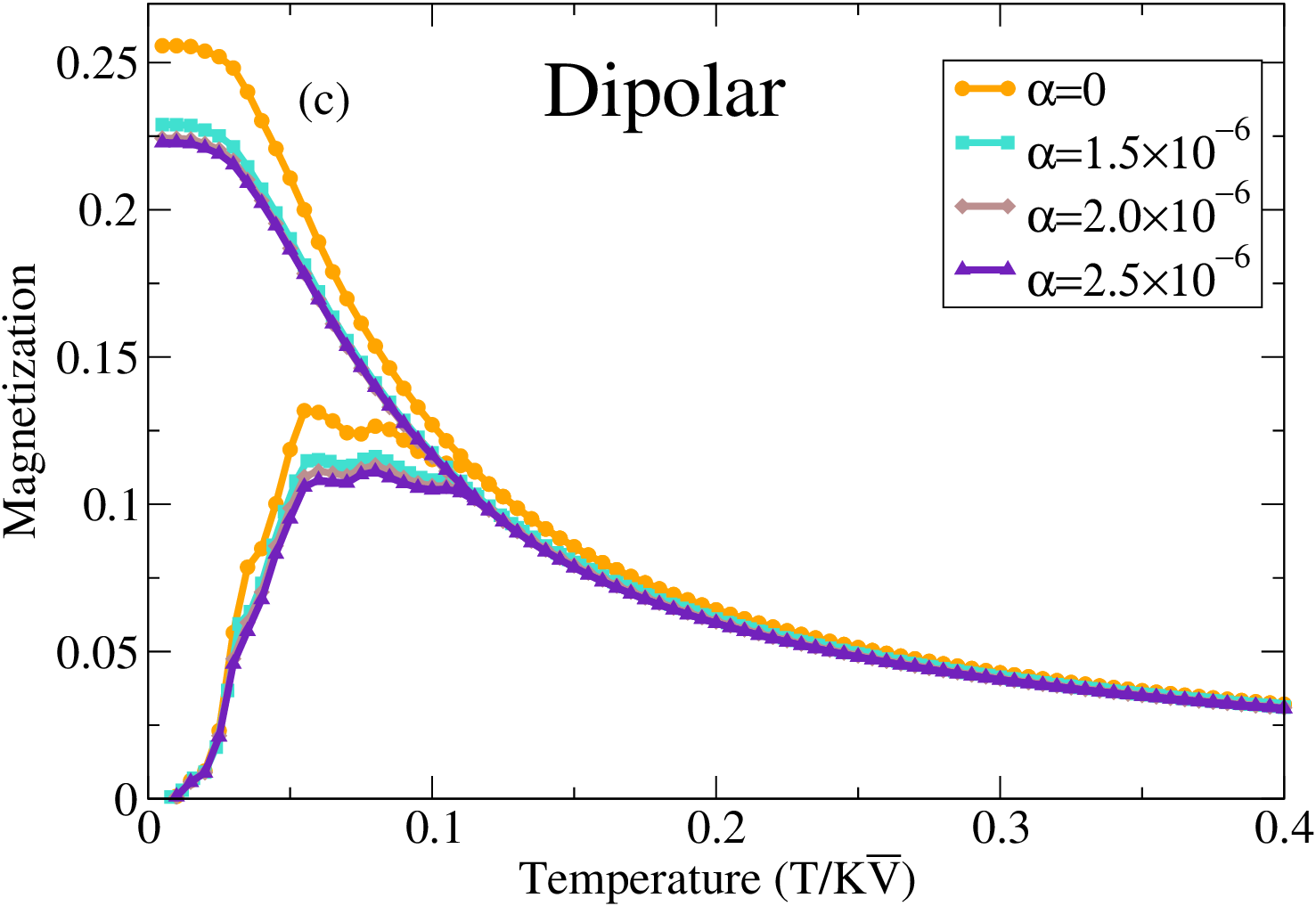,width=1.0\linewidth,clip=} 
\end{tabular}
\caption{ ZFC and FC curves for the (a) NNSR, (b) LRMF and (c) dipolar cases. For each case, there is a constant temperature decrease/increase of $\Delta T=0.004$ for every $100$ s and a field  $h=0.01 $.}
\label{fczfc_int_calc}
\end{figure}
The systems of magnetic nanoparticles are in general polydisperse. The shape and size of the particles are not well-known but the particle-size distribution is often found to be lognormal. \cite{buhrman}
We consider the system consisting of lognormally-distributed and the volume $V_i$ of each particle is obtained from a lognormal distribution 
\begin{eqnarray}
P(V_i) = \frac{1}{\sigma V_i \sqrt{2\pi}} exp \left[  \frac{-(ln(V_i)-\upsilon)^2}{2\sigma^2}\right],
%P(V_i)	= \frac{1}{\sigma V_i \sqrt{2\pi}} e^\frac{-(ln(V_i)-\mu)^2}{(2\sigma^2)}.
\label{dist}
\end{eqnarray}
where $\upsilon=ln(\bar{V})$, $\bar{V}$ is the mean size and $\sigma$ the width of the distribution. The distribution consists of $1000$ particles of sizes between $R =1.3a_0$ and $R=5.3a_0$, where $a_0(=4.17\AA)$ is the lattice parameter of NiO. \cite{hutchings} %A single peak size distribution 
The total number of particles is deliberately chosen to be small in order to simplify the numerical calculation, as well as to get some insight into the role of size-dependent fluctuations in the magnetization in the present study. %A single peak size 
 In all the cases, we use $\sigma=0.6$ and $\upsilon=3.57$. For the dipolar case, the nanoparticles are arranged in a $10\times10\times10$ simple-cubic box. %In this paper we use $\sigma=.6$ and $\upsilon=3.57$. This distribution has been shown in FIG. \ref{fig1}.
\begin{figure}
%\vspace*{1.2cm}
\centering
\includegraphics [angle=0,width=1.0\columnwidth] {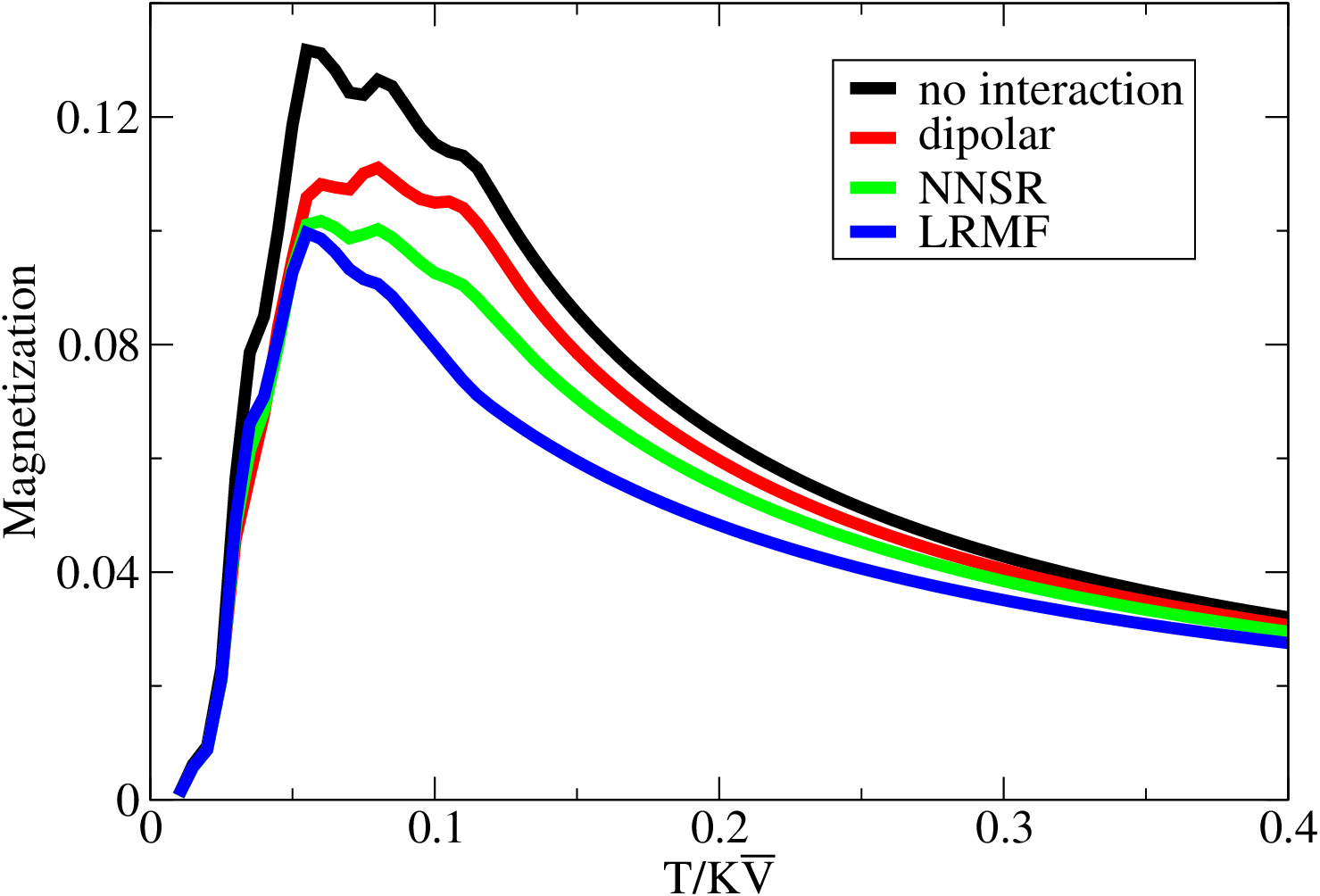}
\caption{A comparison of the ZFC curves for all the interacting cases. Here, again a constant temperature decrease/increase of $\Delta T=0.004$ for every $100$ s and a field of $h=0.01 $. The values of interaction parameters are $\alpha=2.5\times10^{-6}$, $J^{\rm sr}=2.5\times10^{-6}$ and $J^{\rm lr}=4.5\times10^{-8}$ for dipolar, NNSR and LRMF, respectively.}
\label{zfccmp}
\end{figure}
We can solve Eqs. (\ref{mmu}) and (\ref{mtvpv_int}) for any heating/cooling process using any of the interactions discussed above. A recipe to solve these equations for a zero-field-cooled (ZFC) protocol is as follows. The system is cooled from a very high temperature to the lowest temperature in the absence of any magnetic field. The absence of the field during cooling causes a complete demagnetization of the nanoparticles. This condition is the same as  $p_1(0)=1/2$, or $\hat{\mathcal{M}}(0)=0$ in Eq. (\ref{mmu}). Now, a constant field is applied and the system is heated upto a high temperature. At each temperature change, we evolve the system using Eq. (\ref{mmu}). Suppose we increase the temperature from $T$ to $T+dT$. Then, the magnetization at $T+dT$ is given as:
\begin{eqnarray}
 \hat{\mathcal{M}}_{\rm ZFC}(t, T+dT)&=&e^{\mathcal{A}(T+dT)t}\hat{\mathcal{M}}_{\rm ZFC}(0, T+dT) \nonumber \\
&+&(1-e^{\mathcal{A}(T+dT)t})\hat{\nu}(T+dT),\nonumber \\
\label{mmu_zfc}
\end{eqnarray} 
where the initial value of the magnetization vector at $T+dT$ is the same as the magnetization vector at $T$, relaxed for the wait-time $t_w$. Thus, we can write:
\begin{eqnarray}
 \hat{\mathcal{M}}_{\rm ZFC}(0, T+dT)=\hat{\mathcal{M}}_{\rm ZFC}(t_w, T).
\label{m0_zfc}
\end{eqnarray}
Since the elements of the matrix $\mathcal{A}$ and the vector $\hat{\nu}$ depend on $T$, we have shown the functional dependence of $\hat{\mathcal{M}}$ on $T$, which varies during the process. In the above expression, $t_w$ is the wait-time at each temperature-change.  
The magnetization is averaged over the volume distribution $P(V_i)$ as:
\begin{eqnarray}
 %\begin{eqnarray}
 \bar{M}_{\rm ZFC}=\int {M}_{\rm ZFC} (t, T;V_i)P(V_i) d V_i,
\label{mtvpv_int_zfc}
\end{eqnarray}
where $M_{\rm ZFC} (t, T;V_i)$ is the $i^{\rm th}$ element of $\hat{\mathcal{M}}_{\rm ZFC}(t, T-dT)$, given by Eq. (\ref{mmu_zfc}).
We can define the heating rate of the process as the total time elapsed at each temperature-change. Thus, a heating rate of $100$ s per temperature-unit corresponds to a heating process in which the system is relaxed for $t=100$ s at each temperature-step.

Henceforth in this paper, volume $V_i$ is used in units of the average volume $\bar{V}$, which is $193 a_0^3$ in our case. Also, the average anisotropic-energy $\mathcal{K}\bar{V}$ is taken as the unit of energy. By setting $k_{\rm{B}}=1$, we can use $\mathcal{K}\bar{V}$ as a unit of temperature $T$ and the field $\mathcal{H}$. Hereafter, we use a dimensionless quantity $h={\mu}_B H/\mathcal{K} a_0^3$ as the unit of the field, e.g., $h=0.01$ is equivalent to a magnetic field of $300$ Gauss.
%%%%%%%%%%%%%%%%%%%%%%%%%%%%%%%%%%%%%%%%%%%%%%%%%%%%%%%%%%%%%%%%%%%%
\section{ZFC and FC magnetizations }
\label{seczfc_int} 
%%%%%%%%%%%%%%%%%%%%%%%%%%%%%%%%%%%%%%%%%%%%%%%%%%%%%%%%%%%%%%%%%%%%
we begin our study with the most fundamental protocols, i.e., the study of zero-field-cooled (ZFC) and field-cooled (FC) magnetizations.
In the ZFC process, the system is first demagnetized at a very high temperature and then cooled down to a low temperature in a zero magnetic-field. A small
magnetic-field is then applied and the magnetization
is calculated as a function of the temperature. During this heating process, the evolution of magnetization is given by Eqs (\ref{mmu_zfc}) \& (\ref{m0_zfc}). On the other hand in the FC protocol, the system is cooled
in the presence of a small magnetic-field from higher temperatures to a low temperature. For a decrease of the temperature from $T$ to $T-dT$, we can write: 
%In the prsent case we study the effect of various interaction  
\begin{eqnarray}
 \hat{\mathcal{M}}_{\rm FC}(t, T-dT)&=&e^{\mathcal{A}(T-dT)t}\hat{\mathcal{M}}_{\rm FC}(0, T-dT)\nonumber \\
&+&(1-e^{\mathcal{A}(T-dT)t})\hat{\nu}(T-dT),\nonumber \\
\label{mmu_fc}
\end{eqnarray} 
where
\begin{eqnarray}
 \hat{\mathcal{M}}_{\rm FC}(0, T-dT)=\hat{\mathcal{M}}_{\rm FC}(t_w, T).
\label{m0_fc}
\end{eqnarray}
The magnetization is averaged over the volume distribution $P(V_i)$, given as,
\begin{eqnarray}
 %\begin{eqnarray}
 \bar{M}_{\rm FC}=\int {M}_{\rm FC} (t, T;V_i)P(V_i) d V_i,
\label{mtvpv_int_fc}
\end{eqnarray}
where $M_{FC} (t, T;V_i)$ is the $i^{\rm th}$ element of $\hat{\mathcal{M}}_{\rm FC}(t, T-dT)$, given by Eq. (\ref{mmu_fc}).
In the present study, we calculate the ZFC and FC magnetization for various temperature, using long-range and short-range interactions. For all these cases, we use a constant temperature decrease/increase of $\Delta T=0.004$ for every $100$ s. 
In FIG. \ref{fczfc_int_calc} (a), (b) \& (c), we have shown  ZFC and FC  magnetization against temperature plots for NNSR, LRMF and dipolar interactions. Since the interaction plays a role in decreasing the net magnetization, \cite{garcia} we have invoked an antiferromagnetic interaction in the case of LRMF and NNSR interactions. We find that on increasing the temperature, the magnetization $M_{ZFC}$ first increases, attains a maximum at a blocking temperature, and then starts decreasing. 

In all the cases, we find that as we increase the interaction strength, the peak of the ZFC magnetization shifts toward  lower temperatures. In the case of LRMF, we find a dramatic smoothness in the ZFC magnetization as the interaction is increased. The smoothness of the ZFC magnetization indicates a lowering of the effects of size-dependent fluctuations in the magnetization. The reason for the smoothness may be the onset of a collective behaviour of the nanoparticles with the increase in the interaction parameter. An increase in the interaction strength is equivalent to bringing nanoparticles closer to each other. As the nanoparticles come closer to each other, their dynamics loses their individuality and the net effect of other nanoparticles leads to a collective behaviour. The same behaviour can be seen in all the other cases. For the sake of comparison, we have also shown the ZFC magnetization for the non-interacting case as well as for the interacting cases together in FIG. \ref{zfccmp}.
We find that in all the cases in FIG. \ref{fczfc_int_calc}, the FC magnetization $M_{\rm{FC}}$ coincides with 
$M_{\rm{ZFC}}$ at higher temperatures, but departs from the ZFC curve at lower temperatures that are well above the blocking temperature. On further lowering the temperature, the  FC magnetization 
tends to a constant value. The blocking temperature shows a substantial dependence on the heating 
rate. For an infinitely-slow heating-rate, $T_{\rm{B}}$ approaches zero and the ZFC curve shows a similar behaviour as the FC curve. We also find that for a very weak interaction, the FC magnetization never decreases as the temperature is lowered.
But, the increase in the interaction leads to a flatness in the FC magnetization below the blocking temperature which can be seen in all the cases shown in FIG. \ref{fczfc_int_calc}. This flatness in the FC-curves again shows the signature of the co-operative phenomenon of the nanoparticles, arising due to the interactions. 
%For a single peak distribution of smaller sizes ($dist. 1$), as shown in FIG. \ref{fczfc_int_calc}(a) ,
 We see that the FC and ZFC magnetizations in the non-interacting case are greater than those in the interacting cases. We also find that the blocking temperature for the non-interacting case ($T_B=0.04$) is much lower than the corresponding interacting one ($T_B=0.05$). 
\section{AC-susceptibility }
\label{secac}
The relaxation phenomena of a magnetic system are often investigated in the presence of an oscillating magnetic-field. \cite{jonsson1,djuberg,andersson,nordblad1,kleeman1} It is worthwhile to use an oscillatory magnetic-field in our study as the frequency of oscillations of the magnetic field may compare with the inverse of the relaxation time of the nanoparticles. Hence, by changing the frequency of the magnetic field, we can study the dynamics of an assembly of nanoparticles.
For a time varying magnetic field $h(t)=h_0 e^{-i\omega t}$, the analogue of Eq. (\ref{mateq}) is given by:
\begin{eqnarray}
  \frac{d}{dt}\hat{\mathcal{M}} = \mathcal{A}\hat{\mathcal{M}}-\hat{a}(t), 
\label{mateq1}
\end{eqnarray}
where $\hat{a}(t)=\hat{a}_0 e^{-i\omega t}$, and $\omega$ is the frequency of oscillation of the magnetic field. 
Multiplying by $e^{-\mathcal{A}t}$ on both the sides of the above equation Eq. (\ref {mateq1}) gives:
\begin{eqnarray}
 e^{-\mathcal{A}t} \left( \frac{d\hat {\mathcal{M}}}{dt}-\mathcal{A} \hat{\mathcal{M}} \right)=-e^{-\mathcal{A}t} \hat{a}(t),
\end{eqnarray}
which can be written in a simpler form as:
\begin{eqnarray}
 \frac{d}{dt}\left(e^{-\mathcal{A}t} \hat {\mathcal{M}}\right)= -e^{-\mathcal{A}t} \hat{a}(t)
\end{eqnarray}
The solution of the above differential equation is given by,
\begin{eqnarray}
   \hat{\mathcal{M}}= e^{\mathcal{A}t}\hat{C}-\int_0^t e^{\mathcal{A}(t-s)} \hat{a}(s) ds,
\end{eqnarray}
 where $\hat{C}=\hat{\mathcal{M}}_0$. Using $\hat{a}(s)=\hat{a}_0 e^{-i\omega s}$ in the above equation, we get, 
\begin{eqnarray}
  \hat{\mathcal{M}} = e^{\mathcal{A}t}\hat{\mathcal{M}}_0-\int_0^t e^{\mathcal{A}(t-s)} \hat{a}_0 e^{-i\omega s} ds.
\label{matsol1}
\end{eqnarray}
Carrying out the integral in the above equation, we have,
\begin{eqnarray}
  \hat{\mathcal{M}} &=& e^{\mathcal{A}t}\hat{\mathcal{M}}_0+(\mathcal{A}+i\omega \mathcal{I})^{-1}\hat{a}_0 e^{-i\omega t}\nonumber \\
&-&  e^{\mathcal{A}t}(\mathcal{A}+i\omega \mathcal{I})^{-1}\hat{a}_0,
\label{matcom1}
\end{eqnarray}
\begin{figure}[t]
\centering
\begin{tabular}{ccc}
\epsfig{file=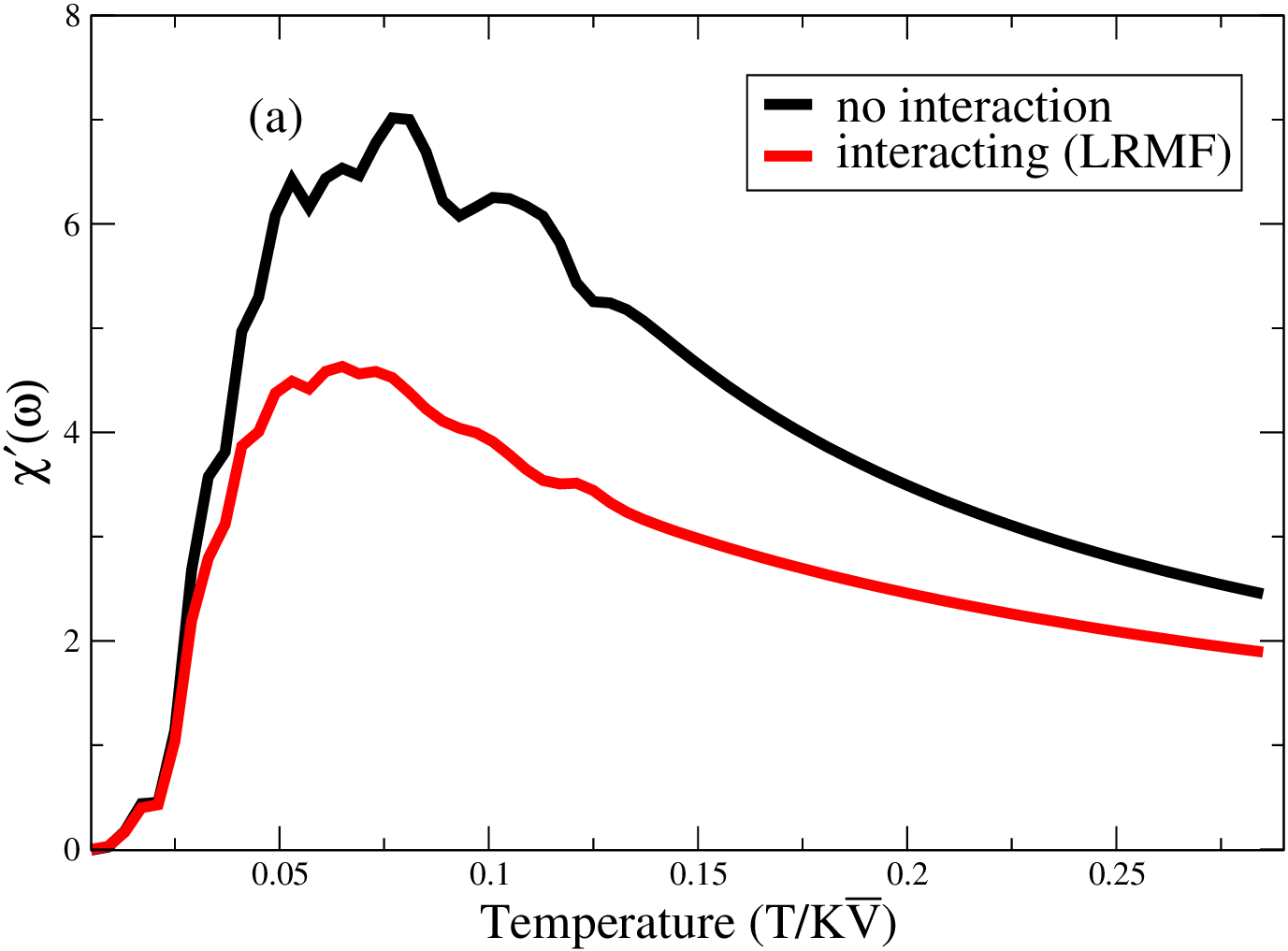,width=1.0\linewidth,clip=} \\
\epsfig{file=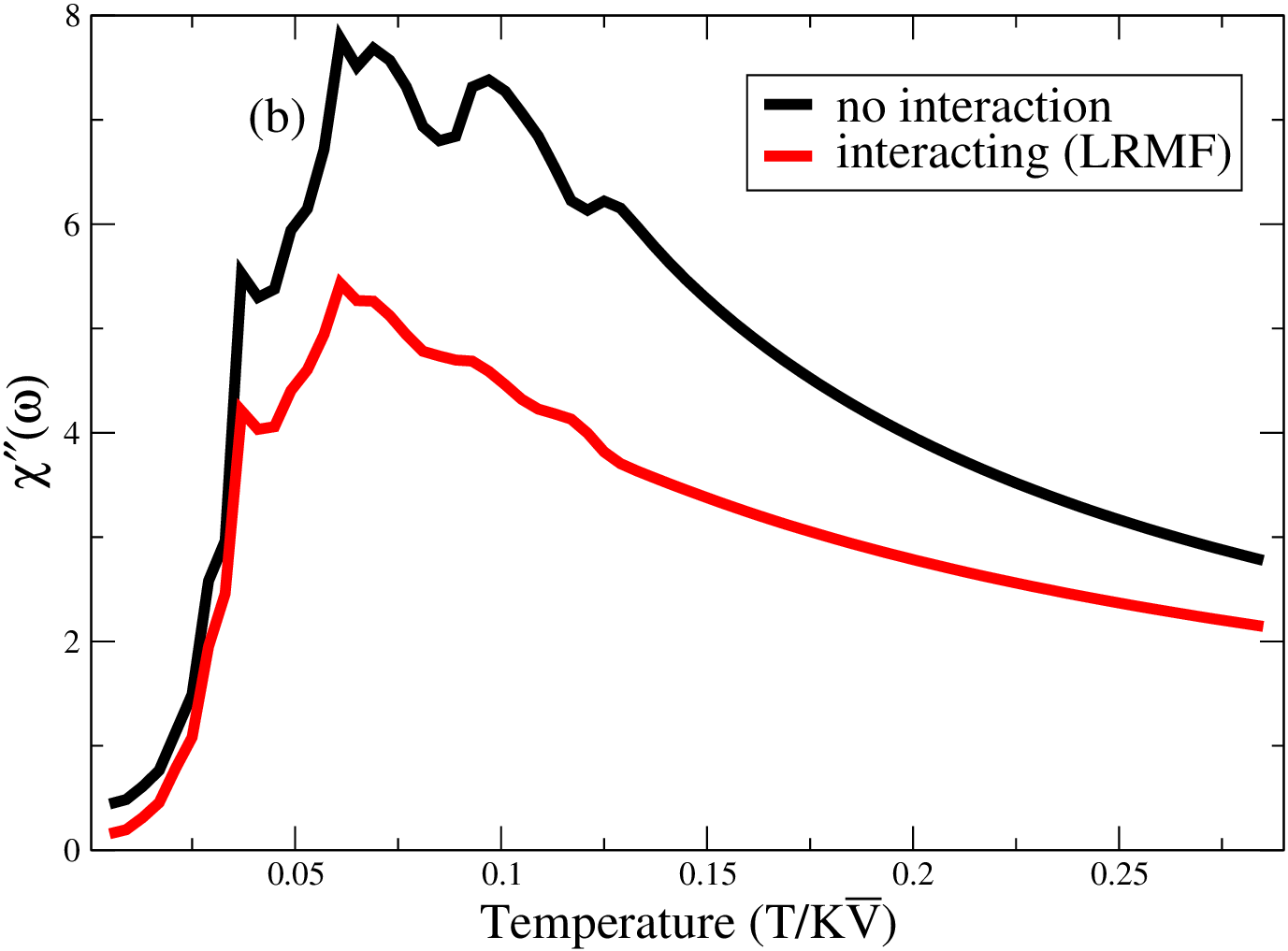,width=1.0\linewidth,clip=} 
\end{tabular}
\caption{A comparison of the (a) real, and (b) imaginary components of the ac-susceptibility for the non-interacting and interacting cases for both the cases, the frequency of the ac-field is $\omega/2\pi=1/500$.}
\label{sus_cmp}
\end{figure}
 where $\mathcal{I}$ is the identity matrix of order $N\times N$. Eq. (\ref{matcom1}) involves the matrix exponential term  $e^{\mathcal{A}t}$, which can be evaluated using diagonal Pad\'e approximants with a scaling and squaring technique from MATLAB. This procedure has been discussed in Section \ref {mod}. Here, using matrix exponential function {\it expm}, we can calculate $e^{\mathcal{A}t}$ in Eq. (\ref{matcom1}). From Eq. (\ref{mtvpv_int}), we can easily calculate $\bar{M}(t)$ by averaging the above value of the magnetization over a size-distribution.
 The response of the time-varying field $h(t)=h_0 e^{-i\omega t}$ can be expressed as the sum of an
in-phase component, $\chi'$ and an out-of-phase component, $\chi''$, of the susceptibility
$\chi(\omega)=\chi'-i\chi''$. 
The in-phase component with the applied field, $\chi'$, is the lossless component, while the out-of-phase component with the applied field, $\chi''$,
is the lossy component. It is obvious that
                                if the change of the external field is very fast as compared with the relaxation time of the 
 particles $\tau$, $ (\omega \gg \tau^{-1} )$, then the particles cannot follow the field
     variation; hence, the magnetization gradually
      decreases to zero as $\omega$ increases. This term is  
          maximum at $ \omega = \tau^{-1}$ and decreases gradually 
          as $\omega$ increases or decreases from the point $\omega = \tau^{-1}$. If the frequency is much less than than re-orientation of the magnetization (i. e. $\omega\ll\tau^{-1}$, then the magnetization is
        always in equilibrium over the time-scale of the measurement. The dynamic
susceptibility can be defined as:

\begin{figure}[t]
\centering
\begin{tabular}{c}
\epsfig{file=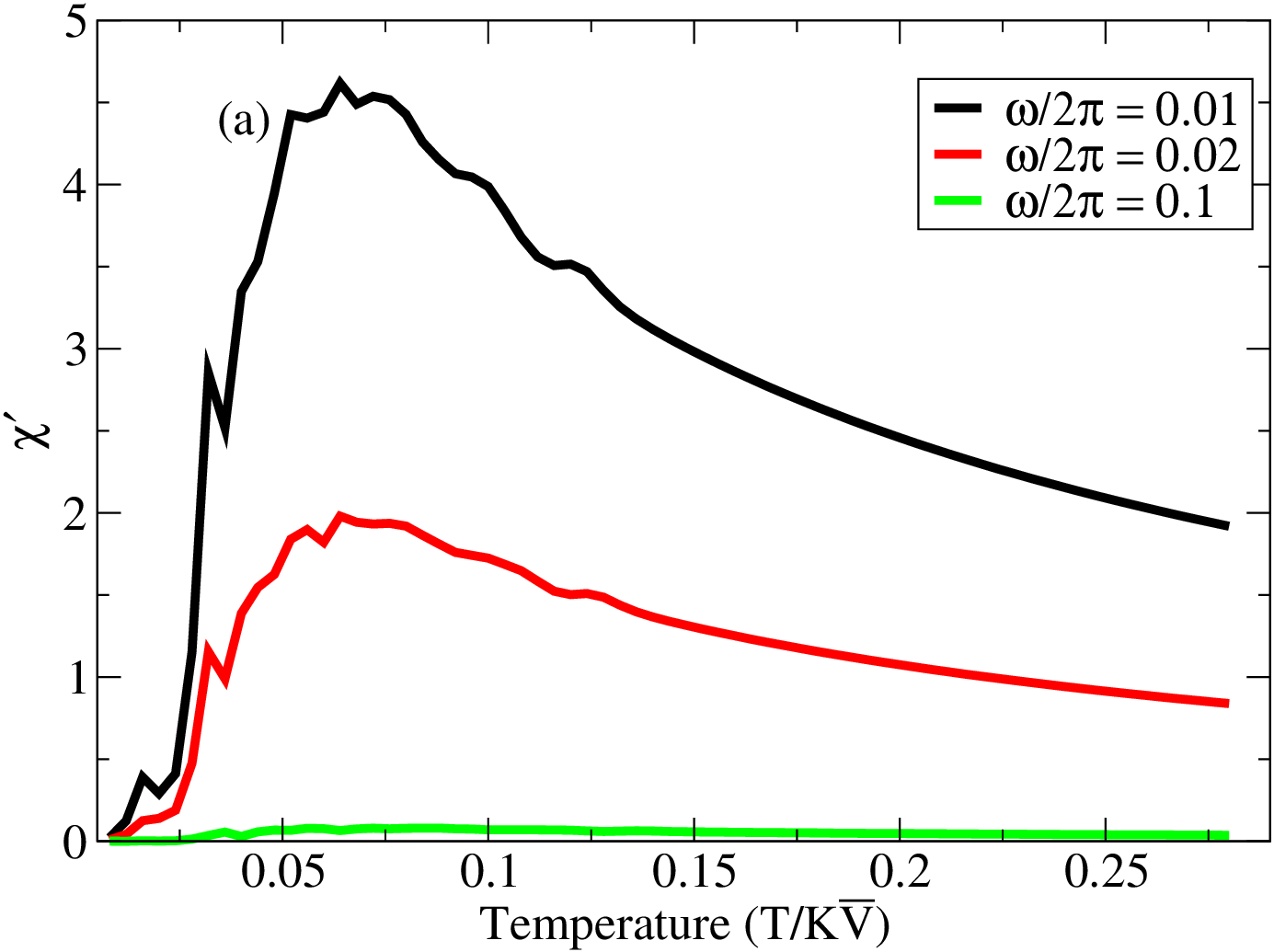,width=1.0\linewidth,clip=} \\
\epsfig{file=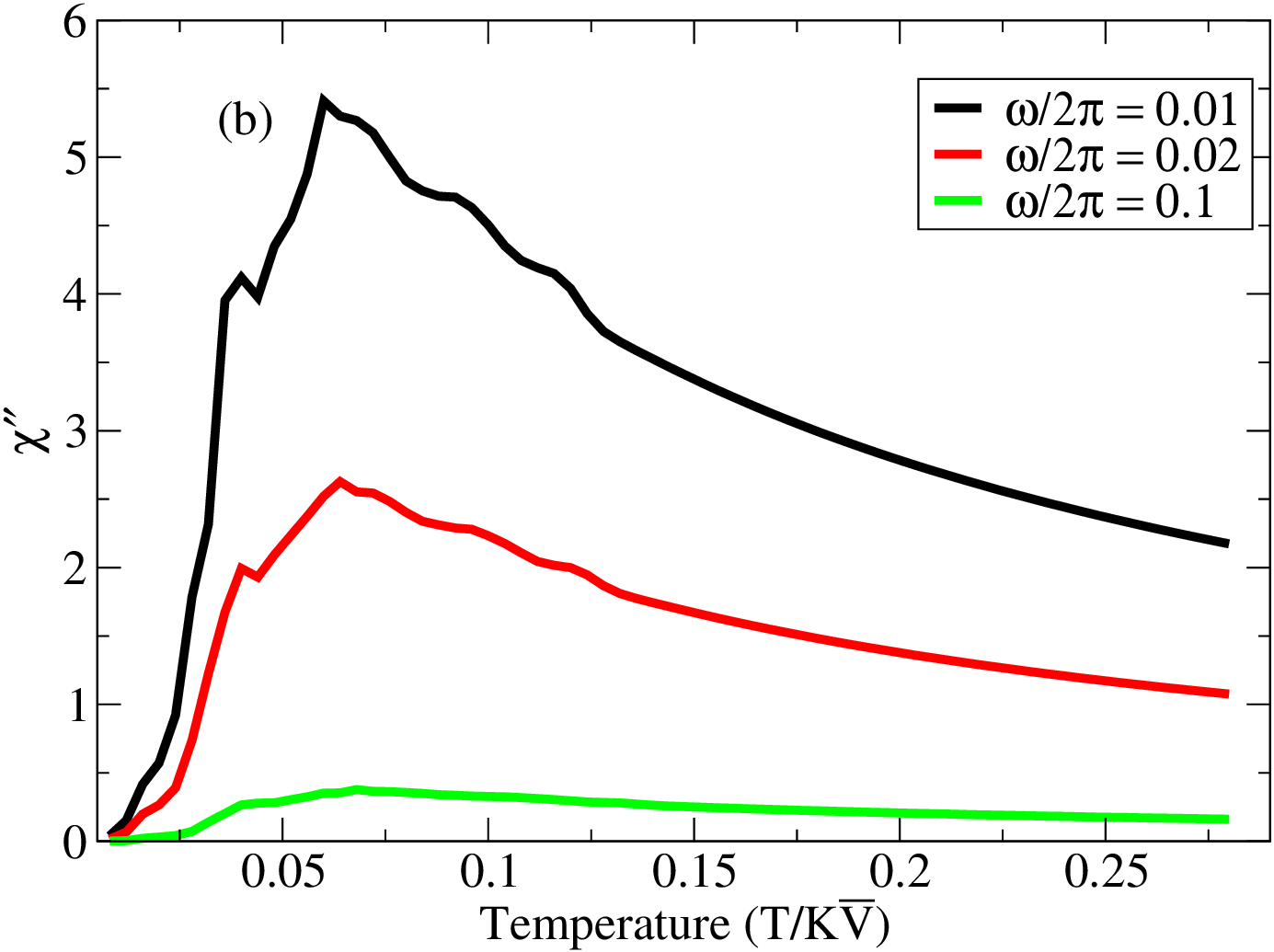,width=1.0\linewidth,clip=} 
\end{tabular}
\caption{ (a) Real and (b) imaginary parts of the susceptibility for the interacting case (LRMF) are plotted against the temperature for various frequencies of the applied ac-magnetic-field.}
\label{susfreq_cmp}
\end{figure}
\begin{eqnarray}
 \chi(t)=M(t)/h(t).
\label{chidef}
\end{eqnarray}
The Fourier transform of $\chi(t)$ can be written as:
\begin{eqnarray}
 \chi(\omega)\equiv \chi'(\omega)-i\chi''(\omega)=\frac{1}{\mathcal{T}}\int_0^{\mathcal{T}}\chi(t) dt.
\label{chifourier}
\end{eqnarray}
Now using $M(t)=M_R+iM_I$, where $M_R$ and $M_I$ are the real and imaginary parts of the magnetization, we can separate $\chi'$ and $\chi ''$, respectively, as:
\begin{eqnarray}
\chi'(\omega) =\frac{1}{h_0\mathcal{T}}\int_0^\mathcal{T}(M_R\rm{sin}(\omega t)+M_I\rm{cos}(\omega t) ),
\end{eqnarray}
and
\begin{eqnarray}
\chi ''(\omega) =\frac{1}{h_0\mathcal{T}}\int_0^\mathcal{T}(M_R\rm{cos}(\omega t)-M_I\rm{sin}(\omega t) ).
\end{eqnarray}
Using the above set of equations, we can calculate the real and imaginary parts of the ac-susceptibility.  For the ac-susceptibility calculations, the system is demagnetized at the lowest temperature and an ac-magnetic-field of amplitude $h_0=0.01$ and frequency $\omega/2 \pi = 0.01$ is applied. The system is then heated upto a high temperature. At each temperature change, we evolve the system for $\mathcal{T}=100$ s and calculate $\chi'$ and $\chi''$ using the above dynamical equations. %We can define heating rate in the process as total time elapsed at each temperature change. Thus heating/cooling rate $10^2$ seconds per temperatue unit corresponds to heating/cooling process in which system is relaxed for $t=100$ seconds at each temperature step.
It has been experimentally observed that ac-susceptibility measurements are sensitive to the interaction effects \cite{djuberg} and confirmed using Monte Carlo simulations. \cite{andersson} We also compare the effect of interactions by considering the non-interacting and interacting cases individually. We find that non-interacting case, which is well described by the individual relaxation of particles differs qualitatively from the interacting case, where the interactions among the particles make the dynamics complex.
The  real and imaginary parts of  ac-susceptibility for the non-interacting case as well as the interacting case using LRMF interactions is shown in  FIG. \ref{sus_cmp}. For both the cases, the frequency of  the ac-field is fixed to $\omega/2 \pi = 1/500$.
  We find that the due to the interaction, the in-phase part as well as the out-of-phase part show reduction in the net value of the susceptibility. 
The effect of size-dependent fluctuations can be observed in the in-phase and out-of-phase components of the susceptibility. The effect of these size-dependent fluctuations are displayed in the non-interacting as well as the interacting cases, as ripples in the ac-susceptibility. But, it is interesting to see that the ripples are smoothened in the interacting cases. The correlated behaviour of nanoparticles is responsible for the averaging of the ripples in this case. The effects of interaction can also be seen as the shifting of the peaks of $\chi'$ and $\chi''$ curves towards lower temperatures.

We have also performed a detailed study of the frequency dependence of the real and imaginary parts of the susceptibility. In FIGs. 
\ref{susfreq_cmp} (a) and (b) we have shown a frequency dependence of the susceptibility for the interacting case by plotting  $\chi'$ and $\chi''$ versus temperature at various frequencies. We see that the increase in the frequency causes a lowering in the magnitude of  $\chi'$ and $\chi''$. Again, we find that the ripples in the susceptibility components due to size-dependent fluctuations get suppressed with an increase in the frequency. We also see that the peak in the $\chi'$ and $\chi''$ curves slightly shifts towards higher temperatures as the frequency is increased. This frequency dependence shows that the particles become less responsive to the magnetic field as the frequency is increased. For a very large frequency the particles may not flip due to the magnetic field which may result in zero-magnetization. 

\section{Memory effects}
\label{secmemory_int} 
%%%%%%%%%%%%%%%%%%%%%%%%%%%%%%%%%%%%%%%%%%%%%%%%%%%%%%%%%%%%%%%%%%%%
\subsection{FC memory-effect}
\begin{figure}[t]
%\vspace{1.4cm}
\includegraphics [angle=0,width=1.0\columnwidth] {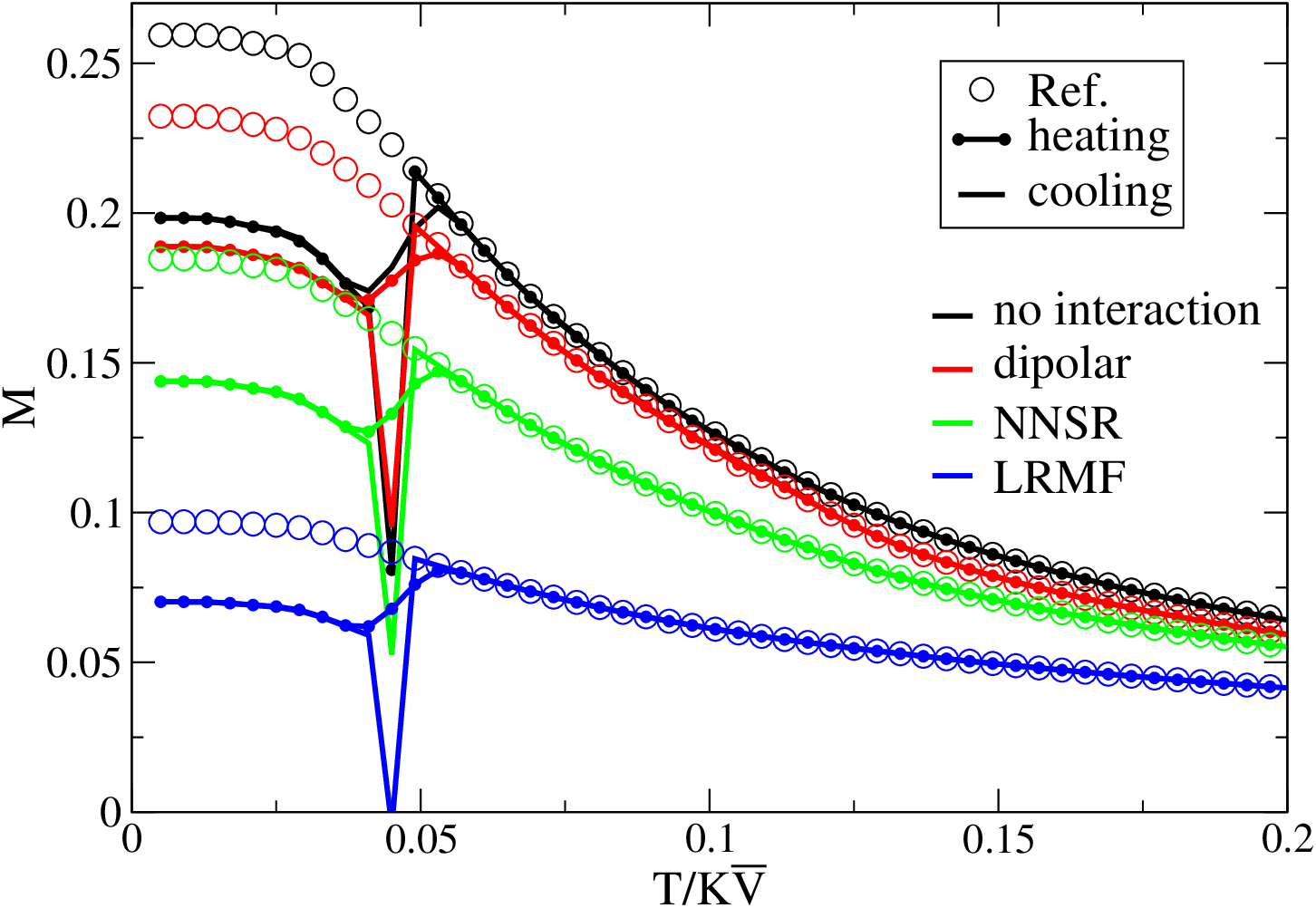}
\caption{FC memory-effects for various interacting cases and the non-interacting case with a stop of $10000$ s at $T=0.04$. The uppermost set of curves shows the non-interacting case, while the dipolar and the NNSR cases follow them towards the lower magnetization. The LRMF case is shown at the bottom.}
\label{fcmem_int}
\end{figure}

Sun {\it et al} \cite{sun} have shown a memory-effect phenomenon in the dc-magnetization
by a series of measurements on a permalloy ${\rm Ni}_{81}{\rm Fe}_{19}$ nanoparticle sample. Later, this memory effect was also been reported by Sasaki {\it et al.} \cite{sasaki} and Tsoi {\it et al.}\cite{tsoi} for the non-interacting or weakly-interacting superparamagnetic system of ferritin nanoparticles and ${\rm Fe}_2{\rm O}_3$ nanoparticles, respectively. All of the studies were concerned only to the ferromagnetic nanoparticles. Experiments on NiO nanoparticles by Bisht and Rajeev \cite{bisht} also confirm a weak memory effect in these particles.
Chakraverty {\it et al.} \cite{malay1} have investigated the effect of polydispersity and interactions among the particles in an assembly of nickel ferrite nanoparticles embedded in a host non-magnetic $\rm{SiO}_2$ matrix. They found that either tuning the interparticle interaction or tailoring the particle-size distribution in a nano-sized magnetic system leads to important  applications in memory devices. In our recent study, \cite{sunil1} we had also shown the memory effects for an assembly of non-interacting NiO antiferromagnetic nanoparticles by the analytical solution of a two-state model. The protocol for the memory effect is as follows. Firstly, we cool the system from a very high temperature to $T_{\rm base}=0.005$ with a probing field of $h=0.01$ switched on. The system is again heated from $T_{\rm base}$ to get the reference curves, which are shown as ``Ref.'' in FIG. \ref{fcmem_int}. The dynamics of the system in this process is the same as that given by Eqs. (\ref{mmu_fc}), (\ref{m0_fc}) \& (\ref{mtvpv_int_fc}) during the cooling process, and Eqs. (\ref{mmu_zfc}), (\ref{m0_zfc}) \& (\ref{mtvpv_int_zfc}), during the heating process.   During these processes, we allow $100$ s to ellapse for every temperature step of $\Delta T=0.001$.
We again cool the system from $T_{\rm H}$ to $T_{\rm base}$ but this time with a stop of $10000$ s at $T=0.04$. The field is cut off during the stop. At the stop, the magnetization is given by,
\begin{eqnarray}
 \hat{\mathcal{M}}_{\rm cool}(t, T_{\rm stop})=e^{\mathcal{A}(T_{\rm stop})t}\hat{\mathcal{M}}_{\rm cool}(0, T_{\rm stop}),
\label{mmu_stop}
\end{eqnarray} 
where
\begin{eqnarray}
 \hat{\mathcal{M}}_{\rm cool}(0, T_{\rm stop})=\hat{\mathcal{M}}_{\rm cool}(t_w, T_{\rm stop}).
\label{m0_stop}
\end{eqnarray}
and
\begin{eqnarray}
 %\begin{eqnarray}
 \bar{M}_{\rm cool}=\int {M}_{\rm cool} (t, T;V_i)P(V_i) d V_i.
\label{mtvpv_int_stop}
\end{eqnarray}
After the pause, the field is again applied and the system is again cooled upto the base temperature $T_{\rm base}$. The recovered magnetization can be given as:
\begin{eqnarray}
 \hat{\mathcal{M}}_{\rm cool}(t, T_{\rm stop}-dT)&=&e^{\mathcal{A}(T_{\rm stop}-dT)t}\hat{\mathcal{M}}_{\rm cool}(0, T_{\rm stop}-dT)\nonumber \\
&+&(1-e^{\mathcal{A}(T_{\rm stop}-dT)t})\hat{\nu}(T_{\rm stop}-dT),\nonumber \\
\label{mmu_stop1}
\end{eqnarray} 
where
\begin{eqnarray}
 \hat{\mathcal{M}}_{\rm cool}(0, T_{\rm stop}-dT)=\hat{\mathcal{M}}_{\rm cool}(t_w, T_{\rm stop}).
\label{m0_stop1}
\end{eqnarray}
The process is shown as the ``cooling'' curves in FIG. \ref{fcmem_int}. Finally, we heat the system at the same rate as that of cooling without any 
stop, which is shown as the ``heating'' curves in FIG. \ref{fcmem_int}. The heating process follows Eqs. (\ref{mmu_zfc}), (\ref{m0_zfc}) \& (\ref{mtvpv_int_zfc}). However, the transition from the cooling process to the heating process at $T_{\rm base}$ is given by,
 \begin{eqnarray}
 \hat{\mathcal{M}}_{\rm heat}(t, T_{\rm base}+dT)&=&e^{\mathcal{A}(T_{\rm base}-dT)t}\hat{\mathcal{M}}_{\rm heat}(0, T_{\rm base}+dT)\nonumber \\
&+&(1-e^{\mathcal{A}(T_{\rm base}-dT)t})\hat{\nu}(T_{\rm base}-dT),\nonumber \\
\label{mmu_stop3}
\end{eqnarray} 
where
\begin{eqnarray}
 \hat{\mathcal{M}}_{\rm heat}(0, T_{\rm base}+dT)=\hat{\mathcal{M}}_{\rm heat}(t_w, T_{\rm base}).
\label{m0_stop3}
\end{eqnarray} 
%The magnetization is averaged over the volume distribution $P(V_i)$ as
%\begin{eqnarray}
 %\begin{eqnarray}
% \bar{M}_{FC}=\int {M}_{FC} (tw,T;V_i)P(V_i) d V_i,
%\label{mtvpv_int_fc}
%\end{eqnarray}
As has been discussed earliar that the interactions among the particles play an important role in the slow dynamics, we study the memory effect using the above protocol for various long-range and short-range interactions. Since the memory-effect phenomenon in the FC magnetization protocol arises  due to polydispersity in the system of nanoparticles, we find a memory dip in all the cases. Yet, we can find out how strong the memory effect is under various interactions. 
 We compare the relative strength of the memory effects for various interactions by introducing a parameter, the Memory Fraction (M.F.), at the stop during the memory measurement, defined as:
\begin{eqnarray}
 {\rm Memory \ Fraction \ (M. F.)}=\frac{\Delta M}{M^{\rm ref}}=\frac{M-M^{\rm ref}}{M^{\rm ref}}
\label{mf_eqn}
\end{eqnarray}
The calculated values of the M.F. for the no interaction, dipolar, NNSR and LRMF cases are given in Table \ref{table1}.
\begin{table}[t]
\centering
\begin{tabular}{|c|c|}
\hline
%\centering
Interaction&M.F.\\

\hline
Non-interacting&$0.24$\\
\hline
Dipolar&$0.18$\\
\hline
NNSR&$0.22$\\
\hline
LRMF&$0.302$\\
\hline
%\label{table1}
 \end{tabular}
%\end{center}
\caption{The memory fraction, defined by Eq. (\ref{mf_eqn}), has been given for the various interactions.}
\label{table1}
\end{table}
  In all the cases, we fix the waiting-time at the stop to be $10000$ s. The trend is a bit surprising. As we know that the memory effect in non-interacting antiferromagnetic nanoparticles is more than that for ferromagnetic particles of the same size-distribution when the distribution consists of smaller sizes. \cite{sunil1} Here, we find that invoking various interactions do change the net memory-dip in antiferromagnetic nanoparticles. As compared to the non-interacting case, we see a reduction in the memory dip for dipolar interactions, an almost similar dip for the  NNSR case, and an increased memory dip for the LRMF case. %We find a decrease in the memory dip for dipolar interaction; it is almost nearer to the non-interacting value for NNSR interactions, and exceeds to the non-interacting value for the LRMF case. 
Since we are employing a simple model in the present study, we find the best effects of the interactions in the case of LRMF. We find that as the interaction parameter increases, the effect of the size-dependent fluctuations decreases, and thus, the memory dip decreases in the dipolar interactions case. On further increasing the interaction parameter, a collective dynamics becomes responsible for the enhanced memory-dip in the NNSR and LRMF cases.

In our earlier work, \cite{sunil1} we had analysed the role of polydispersity in the memory effect for the non-interacting case. As there was no interaction, the dynamics of the individual particles was responsible for this peculiar effect.
However, in the case of interacting antiferromagnetic nanoparticles, we can see that the size-dependent magnetization-fluctuations and interactions among the nanoparticles do modify the memory dip.
This memory effect in an interacting system of nanoparticles can be understood in terms of the droplet picture. \cite{sasaki}
 Droplet theory earlier, proposed for spin-glass systems, is very helpful in getting some insight into spin configurations in these interacting systems of nanoparticles. According to this model for a spin-glass, if the system
is rapidly quenched in a field $h$ to a temperature $T$ below the critical temperature,
spin-glass domains, or clusters called droplets, which are in local equilibrium with respect to $T$, and $h$ grow in size. In this picture, a small temperature-change causes substantial changes in the equilibrium state. At any time $t$, droplets of various sizes exist. Thus, the system is analogous to the non-interacting case where various particle-sizes can be replaced by various clusters grown at temperature $T$ and field $h$.
In droplet theory, \cite{fisher1, fisher2} the dynamics of the droplets is considered to be a thermally-activated process where the energy barrier to form a droplet scales with $L$, the size of the droplet. This droplet picture is relevant to the present study. The thermally-activated dynamics of droplets shows a resemblance to the two-state model
 of the superparamagnet. We see that due to the mean-field interaction with all the other particles, the individual magnetic moment can no longer remains dissociated from the other spins. Hence, an effective moment replaces the individual moment. Since the averaging is done due to the presence of other spins, the net effect on the dynamics is modified according to the interactions between them. 
As the memory dip $\Delta M$ depends strongly on the wait-time parameter and the distribution of sizes the parameter $\Delta M/M $ becomes very effective in the qualitative description of the FC memory-effect.  
 \begin{figure} [t]
% \vspace{1cm}
\centering
 \includegraphics [angle=0,width=1.0\columnwidth] {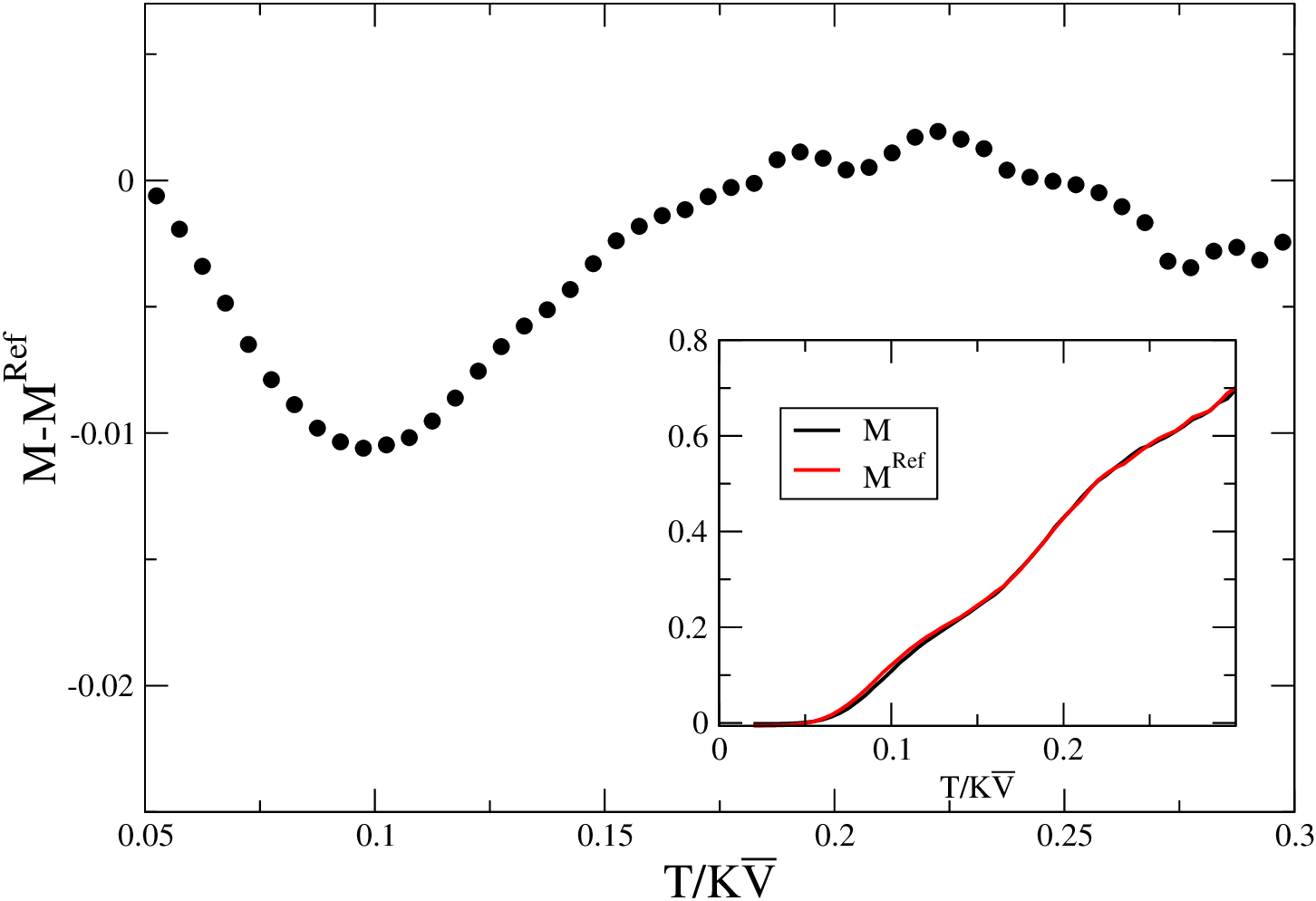}
%\vspace{.3cm}
\caption{ZFC memory-effect using Monte Carlo simulations. The difference between the aged and the normal (reference) ZFC magnetization is shown. The memory dip around the stop temperature $T_s=0.1$ can be seen. In the inset, we have shown the aged and the reference ZFC-magnetizations together.}
 \label {zfcmem}
 \end{figure}
\subsection{ZFC memory effect}
\label{zfceffect}
 Sasaki {\it et al.}\cite{sasaki} suggested a ZFC memory-effect protocol to
confirm whether the observed memory effect is due to glassy
behaviour or not. In this method, we first cool the system rapidly in
a zero-field from a high temperature to the stop
temperature, which is well below the blocking/freezing temperature. The system is left to relax for a wait-time $t_{\rm{w}}$. The rapid cooling
is then resumed down to the lowest temperature where a 
small magnetic field is applied, and the magnetization is calculated during the heating process. A reference ZFC-curve can also be obtained without any stop during the cooling process. If system exhibits memory effect, a memory dip should be observed around the stop temperature during the heating process. Since this dip is substantially smaller, it is better to see the behaviour of the difference between the aged and the normal ZFC magnetizations as a function of the temperature. In the present context of study, we have also investigated the ZFC memory-effect. The system, during sudden cooling under a very small field of $0.000001$ (almost zero-field), is put on hold for a stop of $10^4$ s at $T=0.04$. During heating the system, we find a very small dip at the stop temperature. This very small dip is a drawback of our simplified model.  Since  we did not incorporate randomness in our model, the simple two-state model could not capture the ZFC memory-dip. %case with three distributions of sizes $dist. 1$, $dist. 2$ and $dist. 3$ in FIGs \ref{zfcmem} (a), (b) and (c).
In order to demonstrate the memory effect, we perform Monte Carlo simulations \cite{binder} in a $5\times5\times5$ simulation box of a simple cubic lattice with $125$ particles of different sizes.
 Again, the volume $V_i$ of each particle is obtained from a lognormal distribution as defined in Eq. (\ref{dist}).
In the Monte Carlo method, during each Monte Carlo step (MCS), we select a particle $i$ from $1000$ random sequences of $125$ particles with either `up' or `down' superspin orientations. The attempted flip of the orientation is accepted with a probability of
$\exp(-{E_i}/k_B T)$, where $E_i$ is the energy of the $i^{\rm th}$ particle given by Eq. (\ref{modl}). A Monte Carlo steps is the time-unit in our simulation. Whenever a flip attempt is successful, the magnetic moment of the particle $i$ is updated according to the magnetic-moment versus particle-size plot shown in the inset of FIG. \ref{fig1a}. One Monte Carlo step is equivalent to the intrinsic time-step $\tau_0$ of the superspins. All the processes discussed below using the Monte Carlo method are under fast cooling and heating rates. Though  very fast cooling and heating rates are seemingly far from reality, a qualitative picture of the interacting nanoparticles evolves by this method. 

In the present study, we focus on the dipolar-interaction term and take the interaction parameter to be $\alpha=5\times10^{-7}$. The variation of the interaction parameter $\alpha$  has the same effect as that by changing the lattice constant of the lattice on which the particles are placed.
 We repeat the same protocol for the ZFC memory-effect as discussed above. We first cool the system rapidly in
a zero-field from a high temperature $T=0.6$ to the stop
temperature $T_s=0.1$ (well below the blocking/freezing temperature) and let the system relax for $50000$ MCS. The rapid cooling
is then resumed down to the lowest temperature where a
magnetic field of $h=0.01$ is applied and the magnetization is calculated during the heating process where the temperature is changed in steps of $\Delta T=0.01 K \bar{V}/k_B$ for every  $100$ MCS . A reference ZFC-curve is also obtained without any stop during the cooling process. We have plotted the difference between the aged and the normal ZFC magnetizations as a function of the temperature for the interacting case with dipolar interactions among the particles in FIG. \ref{zfcmem}. For the non-interacting case we pointed out no ZFC memory-effect for a non-interacting assembly of NiO nanoparticles; \cite{sunil1} however, for the present case we see a dip appearing in $\Delta M= M_{\rm ref}-M$ just at the temperature $T_s=0.1$, where the stop and wait has been made during the ZFC rapid cooling process.
\begin{figure} [t]
% \vspace{1cm}
\centering
 \includegraphics [angle=0,width=1.0\columnwidth] {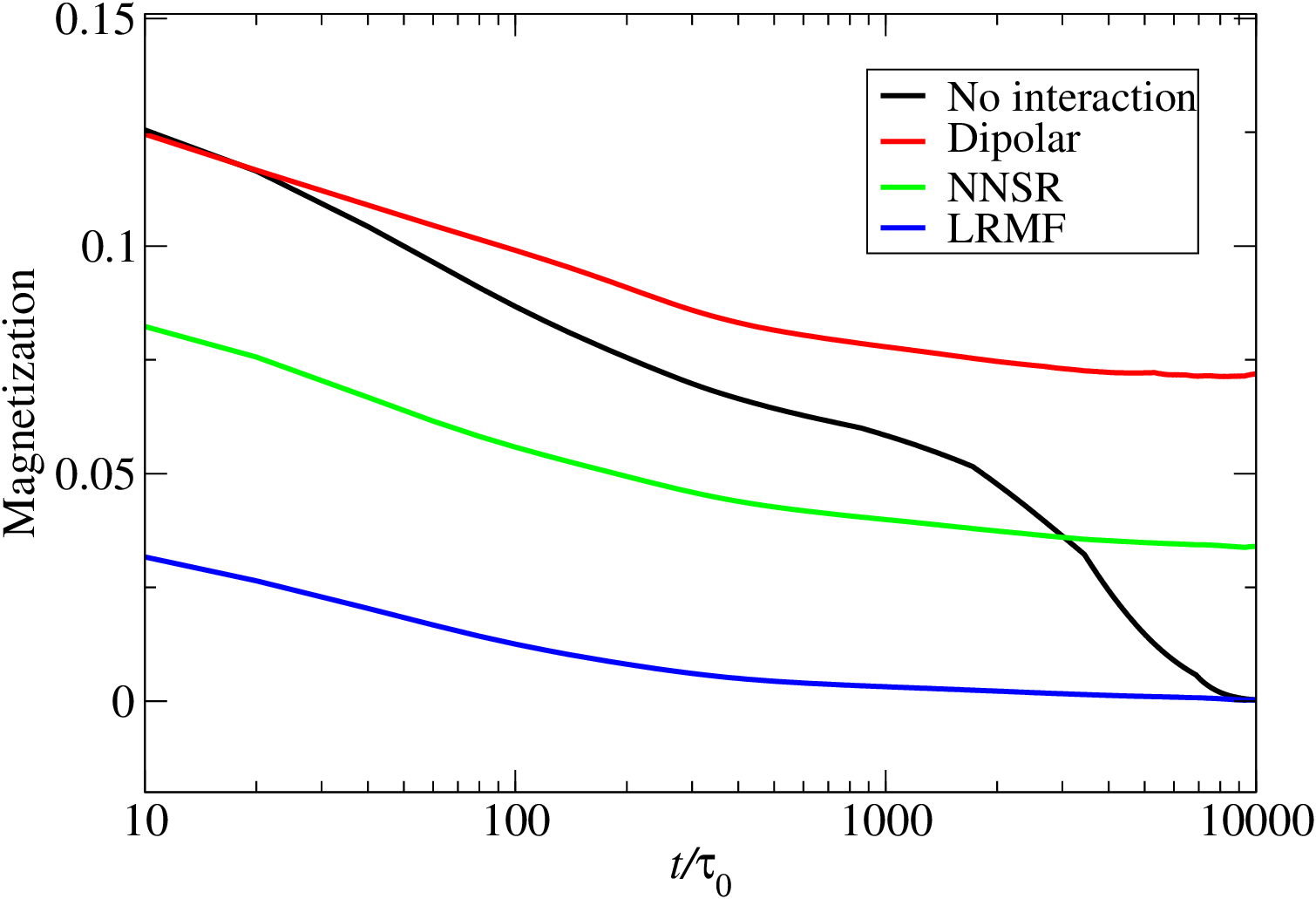}
%\vspace{.3cm}
\caption{A comparison of the aging effect for four cases: No interaction, Dipolar, NNSR, and LRMF. In each case, the wait-time is $t_w=100$ s.}
 \label {int_aging}
 \end{figure}
The ZFC memory-effect can be explained by droplet theory. \cite{sasaki} It has been reported earlier that below a critical temperature droplets (clusters) are growing as the
time elapses. This time-dependent growth of the clusters responsible for the unusual memory-effect in the interacting assembly of nanoparticles. As we allow more time to elapse at $T_s$, the clusters grow proportionally to the wait-time. When the cooling is resumed, this equilibrated clusters freeze at lower temperatures. During the heating process, just near $T_s$, the frozen clusters rearrange themselves, and a dip can be observed in the $\Delta M$ vs. $T$ curve.
\begin{figure*}[t]
 \centering
 \begin{tabular}{cc}
 \epsfig{file=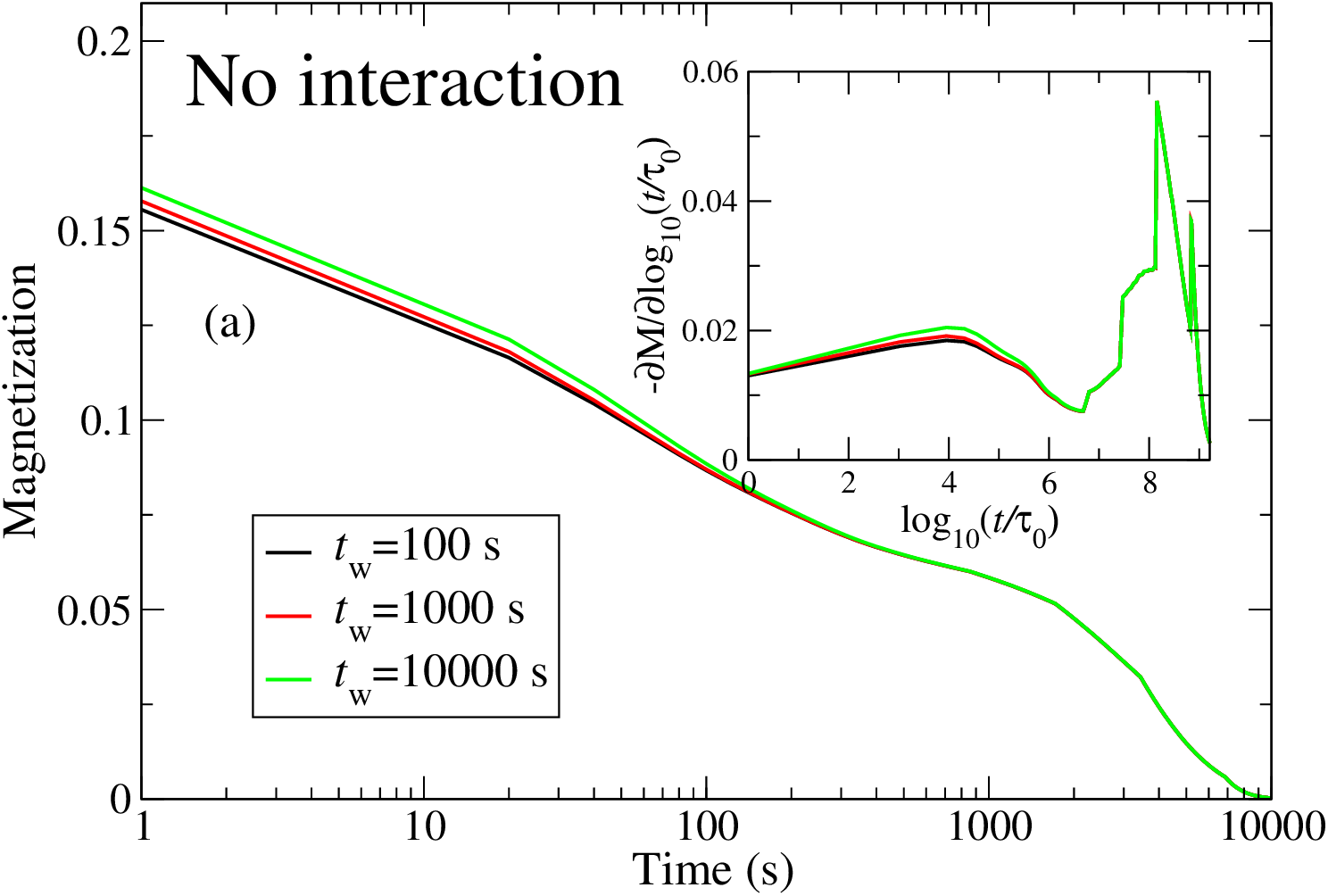,width=.5\linewidth,clip=} &
\epsfig{file=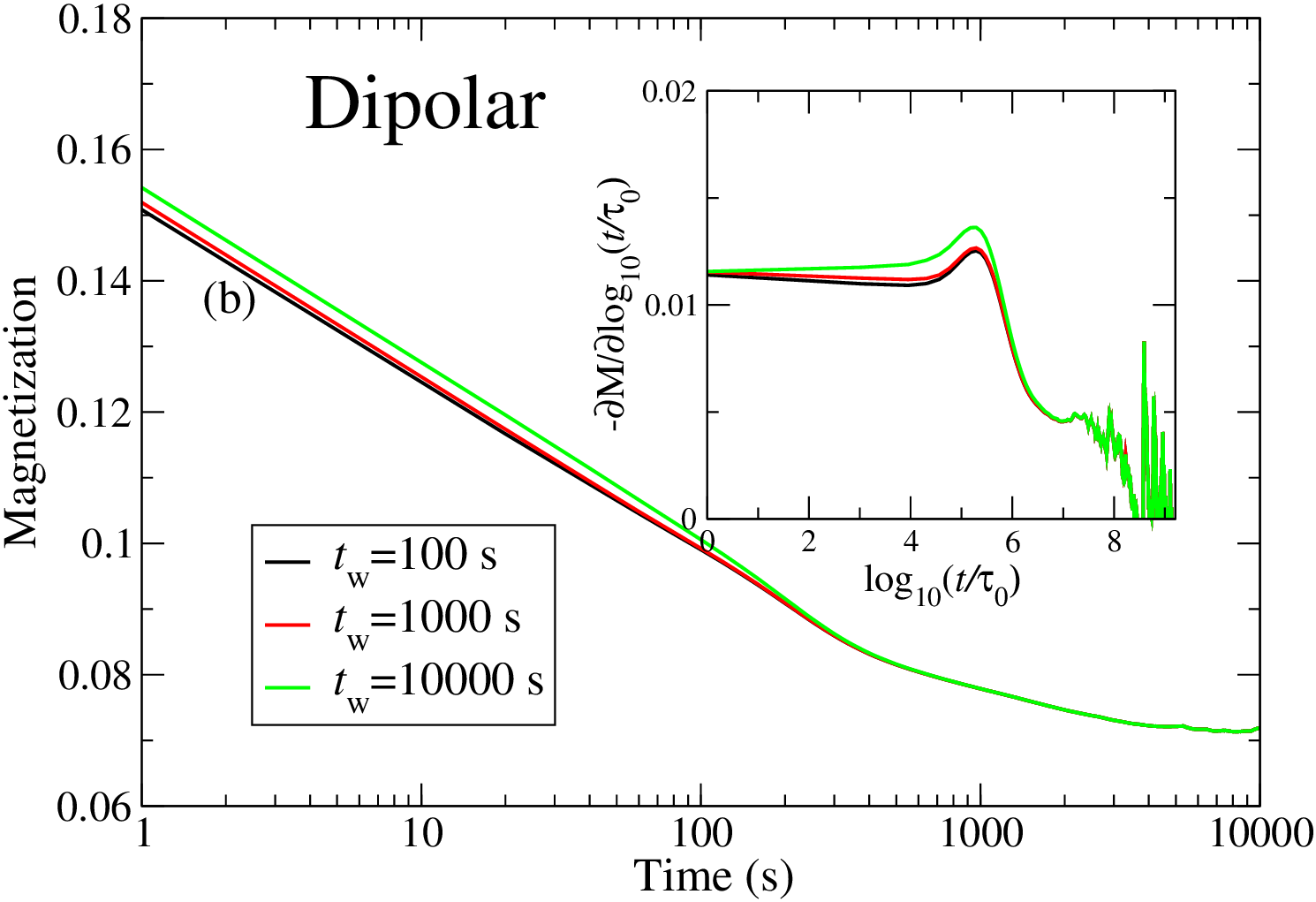,width=.5\linewidth,clip=} \\
 \epsfig{file=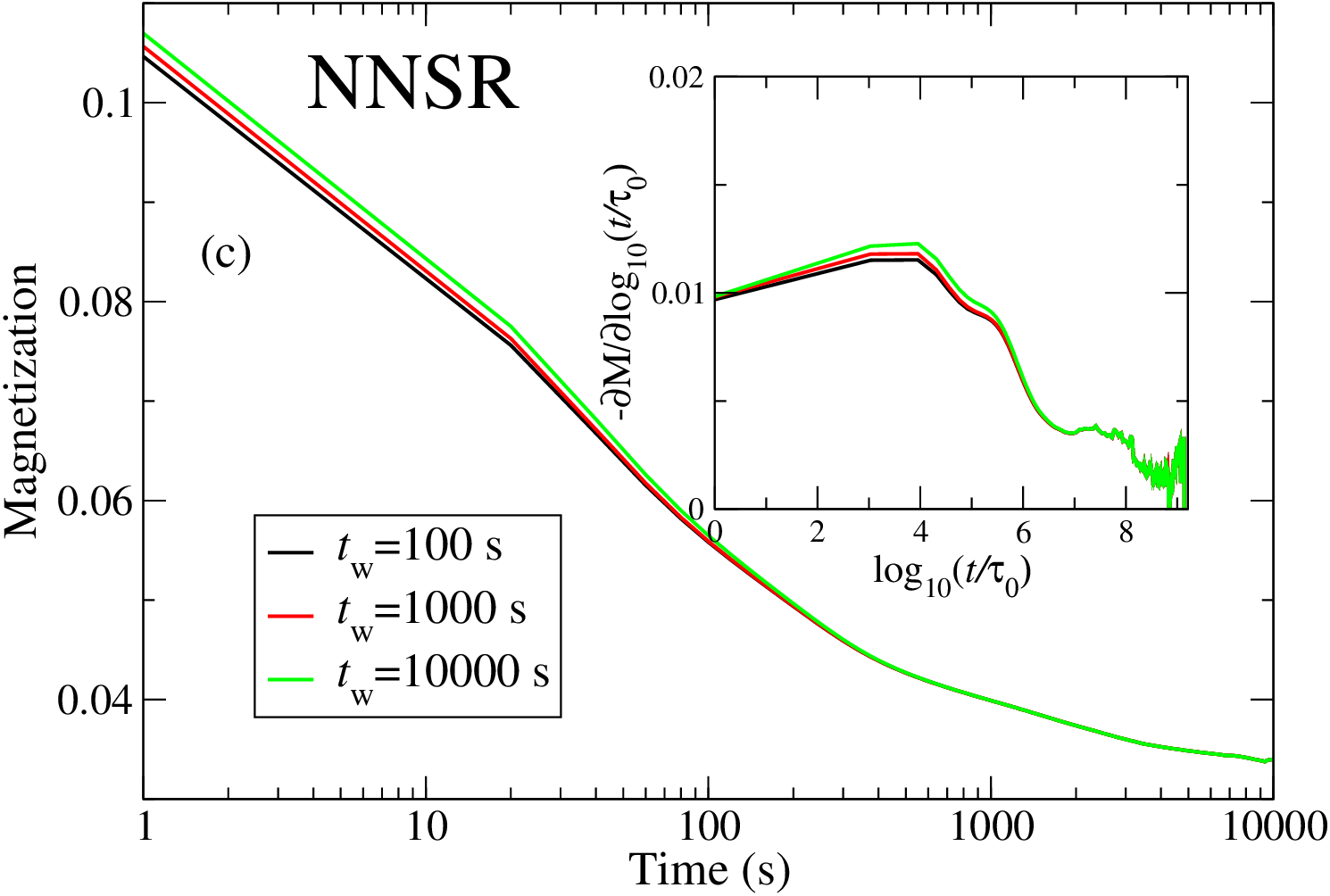,width=.5\linewidth,clip=}&
 \epsfig{file=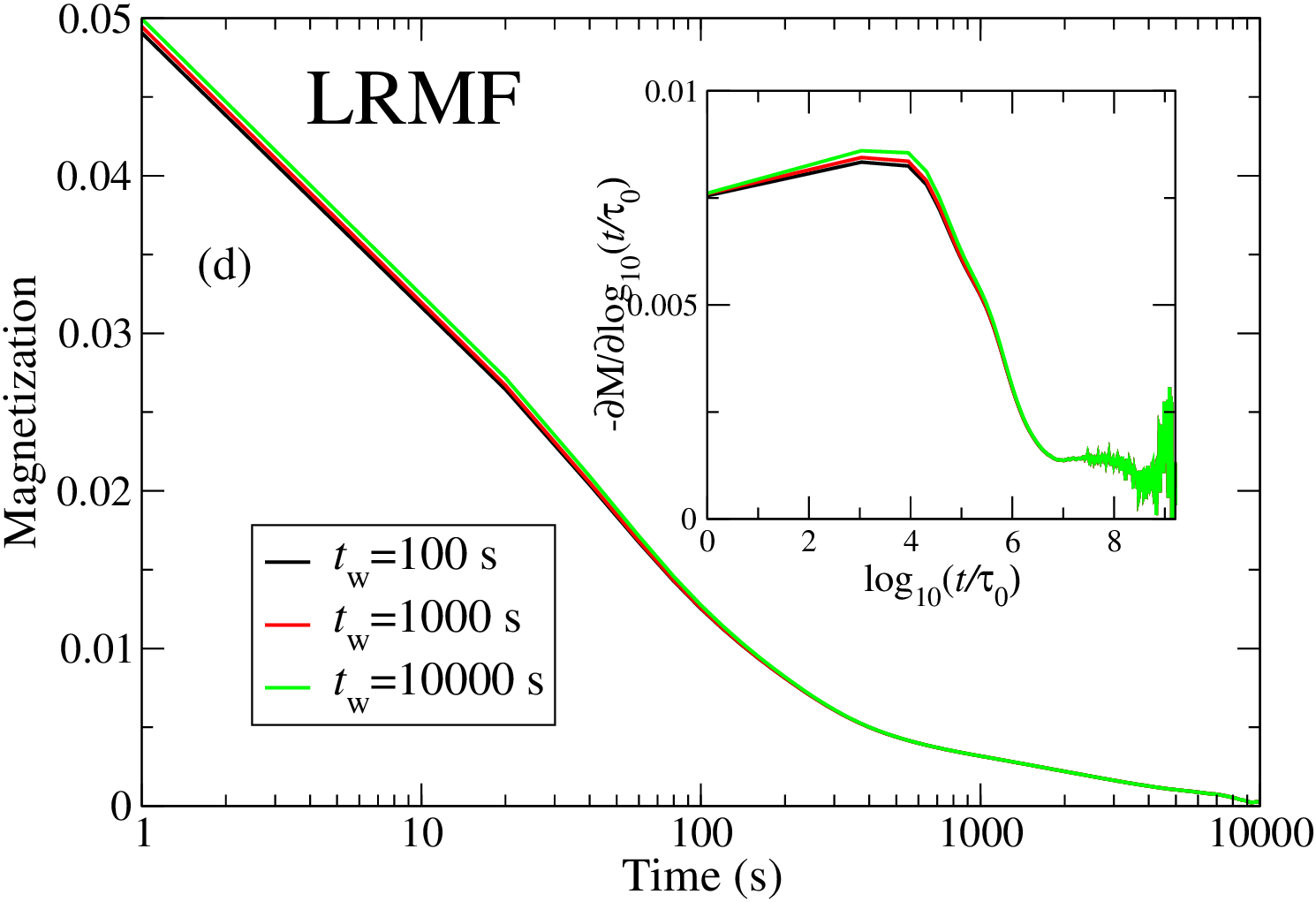,width=.5\linewidth,clip=}
\end{tabular}
 \caption{The aging effect for waiting-times $t_w=100$ s, $1000$ s and $10000$ s has been shown for: (a) no interaction, (b) dipolar c) NNSR, and d) LRMF interactions.  In the interacting cases (b) (c) and (d), we see a smooth decay of the magnetization with time.}
 \label{aging_int_cmp1}
 \end{figure*}
\section{Aging effects}
\label{secaging_int}
The magnetization relaxation of nanoparticles can also be studied by analysing the wait-time dependence of magnetization. In one such protocol, the system is cooled from a high temperature to the lowest temperature in the presence of a constant magnetic-field. During the cooling process, the system is halted at a temperature lower than the blocking temperature. The system is paused for a wait-time $t_w$ before switching off the magnetic field. The relaxation of magnetization show the $t_w$ dependence. This phenomenon is known as aging effect. 
Aging effects in  polydispersed assemblies of interacting nanoparticles have been been receiving a great deal of attention. The aging effects in interacting nanoparticles has been indicated a glassy behaviour in these systems. The aging-effect has been well-studied in the spin-glass systems, \cite{nowak1,lundgren,ocio1} as well as in interacting assemblies of ferromagnetic nanoparticles. \cite{sasaki,sahoo1,sahoo2,sahoo3,tsoi} The aging effect or wait-time dependence in the thermoremanant magnetization (TRM) protocol has also been observed in non-interacting assemblies of nanoparticles. Most of the aging effect studies are centred around ferromagnetic nanoparticles. However, recent experiments on antiferromagnetic NiO nanoparticles show aging effect in these systems also. \cite{bisht} In our earlier work, \cite{sunil1} we had shown the aging effect for a collection of a few non-interacting NiO nanoparticles. We find that in the non-interacting case, the wait-time dependence was due to an ill-defined initial state of the system just before switching off the field. When the interaction among the particles comes into the picture, we should be able to observe the effects arising due to the collective behaviour of the nanoparticles.  Most of the aging experiments are performed by the measurement of the time-dependent ZFC and TRM magnetizations. The methodology in a TRM protocol is as follows. The system is cooled in a field to a base temperature $T_{\rm{base}}$ below the blocking temperature $T_{\rm B}$. During the cooling process, the magnetization is given by Eqs. (\ref{mmu_fc}), (\ref{m0_fc}) \& (\ref{mtvpv_int_fc}). After a waiting-time $t_{\rm{w}}$, the magnetic field is switched off and we find the relaxation in the magnetization. The magnetization in this case is given by, 
\begin{eqnarray}
 \hat{\mathcal{M}}_{\rm TRM}(t, T_{\rm stop})=e^{\mathcal{A}(T_{\rm stop})t}\hat{\mathcal{M}}_{\rm TRM}(0, T_{\rm stop}),
\label{mmu_stop}
\end{eqnarray} 
where
\begin{eqnarray}
 \hat{\mathcal{M}}_{\rm TRM}(0, T_{\rm stop})=\hat{\mathcal{M}}_{\rm TRM}(t_w, T_{\rm stop}),
\label{m0_stop}
\end{eqnarray}
and
\begin{eqnarray}
 %\begin{eqnarray}
 \bar{M}_{\rm TRM}=\int {M}_{\rm TRM} (t, T;V_i)P(V_i) d V_i.
\label{mtvpv_int_stop}
\end{eqnarray}
when the TRM magnetization is plotted against the time, it is found that the relaxation in the magnetization depends upon the waiting-time $t_{\rm{w}}$. 

 In this section we will study the aging effect for a polydispersed system of interacting NiO nanoparticles in the TRM protocol. Our investigation
is carried out by cooling the system in the presence of magnetic field of $h=0.01$ upto the base temperature $T_{\rm{base}}=0.024$ and then cutting the field off after a wait-time $t_{\rm{w}}$ to let the system relax. In a previous work, \cite{sunil1} we had discussed the aging effects for various size-distributions and made a comparative study with the ferromagnetic particles case. However, in this section, we perform a comparative study of the various interactions in NiO nanoparticles, as discussed in Section \ref{mod}. In FIG. \ref{int_aging}, we plot the thermoremanant magnetization against the time on a logarithmic-scale for the non-interacting as well as the interacting cases: dipolar, NNSR and LRMF. Before switching off the field, we wait for $t_w=100$ s. We see that the decay in the magnetization is fastest in the non-interacting case, which shows a two-step relaxation. \cite{sunil1} In the non-interacting case, the dynamics of an individual particle is not correlated with that of others in its surroundings. The role of size-dependent fluctuations can be seen as ripples in the relaxation of the magnetization, but in the interacting cases, we see a smoother curve which shows the averaging of the fluctuations in the magnetization due to the mean field arising from the neighbouring particles. Due to the interactions the magnetization persists for a longer time as compared to the noninteracting case. Hence, we can say that  due to the interactions, the dynamics of the nanoparticles gets slower. We have also plotted the magnetization versus logarithmic-time for various wait-times $t_w=100$ s, $1000$ s, and $10000$ s in FIG. \ref{aging_int_cmp1}. We see a wait-time dependence in all the cases, though weak. In the inset of these curves, we have shown the corresponding relaxation-rate for each case, defined as $-\partial M / \partial \log_{10}(t)$. Multiple peaks in all the cases reflect the distinction between the antiferromagnetic nanoparticles and the ferromagnetic nanoparticles: an effect of size-dependent magnetization-fluctuations. In a nutshell, we conclude that the role of size-dependent fluctuations in the magnetization dynamics has been controlled a little bit due to the interactions. The decay of the thermoremanant magnetization gets slower as the interaction is increased.
\section{Conclusions}
\label{secconclusion_int} 
%%%%%%%%%%%%%%%%%%%%%%%%%%%%%%%%%%%%%%%%%%%%%%%%%%%%%%%%%%%%%%%%%%%%
We have studied the slow dynamics of an interacting assembly of antiferromangetic nanoparticles by solving a two-state model analytically. A collection of a few interacting antiferromagnetic nanoparticles has been comparatively studied using various long-range and short-range interactions. We find that the ZFC magnetization shows ripples in all the cases, which starts decreasing as the interaction parameter increases. 
Due to the interactions, the dynamics of the system is governed not only by 
 a broad distribution of particle relaxation-times arising from the polydispersity, but also by a collective behaviour amongst the particles themselves. We have also studied the dynamics of the system using an ac magnetic-field. We find that the real and imaginary components of the ac-susceptibility show ripples, which are due to the size-dependent magnetization-fluctuations. The effects of these size-dependent fluctuations are displayed in the non-interacting as well as interacting cases; however, we find that the ripples are smoothened in the interacting cases. Due to the interaction, we also find a shifting of the peaks of the $\chi'$ and $\chi''$ curves towards lower temperatures. We have also calculated the frequency dependence of  $\chi'$ and $\chi''$,  and found that the frequency causes a lowering in the magnitudes of $\chi'$ and $\chi''$. The frequency dependence of $\chi'$ and $\chi''$shows that the particles become less responsive to the magnetic field as the frequency is increased. 
% Again we find that the ripples in the susceptibility components due to size-dependent fluctuations get suppressed with increase in the frequency.
 we have also shown the memory-effect in the field cooled magnetization protocol for 
an interacting polydispersed assembly of nanoparticles. We have studied the effects of interactions on the memory dip by varying the interaction parameter. As compared with non-interacting case, we find a reduced memory-dip for dipolar interactions, an almost same memory-dip for NNSR interactions, and an increased memory-dip for LRMF interactions. Due to the limitations in our model, we find the best effect of the interactions only in the case of LRMF. We find that as the interaction parameter increases, the effect of the size-dependent fluctuations decreases, and thus, the memory dip decreases in the dipolar interactions cases. On further increasing the interaction parameter, a collective dynamics becomes responsible for the enhanced memory dip in the NNSR case and LRMF cases.
The memory effect and a weak aging effect have also been discussed for all the cases. For the non-interacting case, the size-dependent fluctuations play a role in the obtaining of a high memory fraction. As the interactions are incorporated, the memory fraction decreases, but for the strong interaction cases, the memory fraction increases.
 As the present model could not incorporate randomness and disorder, we have performed a Monte Carlo study considering the dipolar interactions among the particles.  We find a memory dip in the ZFC magnetization, which indicates the importance of the interactions among the particles. The ZFC memory-effect is attributed only to the collective dynamics of the nanoparticles.
The system of interacting nanoparticles under various interactions also shows a wait-time dependence. The decay of thermoremanant magnetization slows down as the interaction among the nanoparticles increases. We have done a comparative study of the aging effects in the no interaction, dipolar, NNSR and LRMF cases.

\end{document}